\documentclass{aa}  

\usepackage{graphicx,epsfig,amsmath,amssymb,color,,subcaption,stmaryrd,multirow,makecell}
\usepackage{txfonts}
\bibpunct{(}{)}{;}{a}{}{,}
\begin{document}

   \title{Formation of proto-cluster: \\
   a virialized structure from gravo-turbulent collapse}

   \subtitle{I. Simulation of cluster formation in collapsing molecular cloud}

   \author{
          Yueh-Ning Lee \inst{1}
          \and
          Patrick Hennebelle\inst{1,2}
          }

   \institute{Laboratoire AIM, Paris-Saclay, CEA/IRFU/SAp -- CNRS -- Universit\'{e} Paris Diderot, 91191 Gif-sur-Yvette Cedex, France\\
              \email{yueh-ning.lee@cea.fr}
         \and
             LERMA (UMR CNRS 8112), Ecole Normale Sup\'{e}rieure, 75231 Paris Cedex, France\\
             \email{patrick.hennebelle@lra.ens.fr }
             }

   \date{Received 16 December 2015; accepted 25 March 2016}

 
  \abstract
   {
   Stars are often observed to form in clusters and it is therefore important to understand how such a region of 
concentrated mass is assembled out of the diffuse medium.
   The properties of such region eventually prescribe the important physical mechanisms and determine the characteristics of the stellar cluster.
   }
   {We study the formation of a gaseous proto-cluster inside a  molecular cloud
   and associate its internal properties to those of the parent cloud by varying the level of the initial turbulence of the cloud, with a view to better characterize the subsequent stellar cluster formation.}
   {High resolution MHD simulations of gaseous proto-clusters forming in molecular clouds collapsing under self-gravity are performed.
   Ellipsoidal cluster regions are determined using gas kinematics and sink particle distribution,
   allowing to determine the mass, size, and aspect ratio of the cluster.
   The cluster properties such as kinetic and gravitational energy are studied and links are made to the parent cloud.
      }
   {The gaseous proto-cluster is formed out of global collapse of a molecular cloud,
     and has non-negligible rotation due to angular momentum conservation during the collapse of the object.
     Most of the star formation occurs in this region which occupies only a small volume fraction of the whole cloud.
   This dense entity is a result of the interplay between turbulence and gravity.
   We identify such regions in simulations and compare the gas and sink particles to observed star-forming clumps and embedded clusters respectively.
  The gaseous proto-cluster inferred from simulation results present a mass-size relation that is compatible with  observations. 
   We stress that the stellar cluster radius, although clearly correlated with the gas cluster radius, 
   depends sensitively on its definition. 
   Energy analysis is performed to confirm that the gaseous proto-cluster is a product of gravo-turbulent reprocessing and that the support of turbulent and rotational energy against self-gravity yields a state of global virial equilibrium although collapse is occurring at smaller scale and the cluster is forming stars actively.
  This object then serves as the antecedent of the stellar cluster, to which the energy properties are passed on.
}
   {The gaseous proto-cluster properties are determined by the parent cloud out of which it forms,
   while the gas is indeed reprocessed and constitutes a star forming environment different from that of the parent cloud.
      }


   \maketitle


\section{Introduction}
An essential step to the understanding of how and in which conditions stars form is to unveil the process through which the dense prestellar cores collapse from molecular clouds.
In terms of density, this process implies a change of more than 10 orders of magnitude. 
To reduce the complexity, 
the study of star formation is often decomposed into hierarchical steps.
Stars are frequently observed to be born in clusters and  many studies stress on the importance of understanding star formation in such context \citep{Lada03,Longmore14}.
In this work, 
we focus on the formation of gaseous proto-cluster, the primary stage of cluster formation, 
that is to say the phase during which it assembles its gas and converts it into stars, 
with a view to better characterize the star formation environment and to prescribe more precisely the physical mechanisms at play.

The catalog reported by \citet{Lada03} for embedded clusters is consistent with the mass-size relation $R \sim M^{1/3}-M^{1/2}$ for low mass clusters as pointed out by \citet{Murray09}, where $R$ and $M$ are respectively the gas radius and mass. 
While the stellar mass of an embedded cluster is not directly observable and is obtained by assuming an underlying universal initial mass function (IMF),
\citet{Adams06} compiled data from \citet{Lada03} and \citet{Carpenter00} and showed a number-size relation $R_* \sim N_*^{1/2}$ between
the cluster radius and the number of objects it contains.
The results of \citet{Gutermuth09} are also compatible with this relation.
The number-size relation is reasonably equivalent to the mass-size relation if we adopt an universal IMF and thus similar averaged mass among clusters.
Despite the slight uncertainty in the power-law exponent, varying from $1/3$ to $1/2$, 
and the scatter of the data,
it is clear that embedded clusters follow some mass-size relation.
Larger data sets of star-forming clumps, identified as gaseous proto-clusters, 
also exhibit a mass-size relation. 
\citet{Fall10} compiled in their Figure 1 the observations of \citet{Shirley03, Faundez04, Fontani05} and found by least-squares regression the relation $R \propto M^{0.38}$.
A regression fit on the ATLASGAL survey  \citep{Urquhart14} data gives a dependence of $R \propto M^{0.50}$.
Note that in their work they fitted $\log (M)$ to $\log (R)$ and found $M \propto R^{1.67}$, 
of which the power-law exponent is not the inverse of what we inferred. 
This indicates that the clump properties are differently dispersed in size and mass, 
and thus there exist some uncertainties in the power-law dependence.
Meanwhile, both studies exhibit mass scatter of about 1 dex and are compatible with constant gas surface density, 
that is, $R \propto M^{0.5}$.
These observations are done with molecular lines and dust continuum of the star-forming gas,
suggesting that the mass-size relation and probably some other properties 
of the stellar cluster are established as early as at the gas-dominated phase.
\citet{Pfalzner16} pointed out that the mass-size relation for embedded clusters and gaseous proto-clusters follow the same trend with different absolute value, which could be explained with star-formation efficiency or cluster expansion. 
One interesting question to ask would be what physical processes actually determine this mass-size relation.
Its existence suggests that this primary phase of cluster formation, 
the gaseous proto-cluster,
is very likely in some equilibrium state governed by the molecular cloud environment in which it resides,
and it is crucial for understanding the nature of the cluster and, more generally, star formation.
An analytical study by \citet{Hennebelle12} yielded, by linking the gaseous proto-cluster to properties of the parent cloud, 
$R \sim M^{1/3}$ or $R \sim M^{1/2}$ for proto-clusters with different accretion schemes. 
We stress that we are emphasizing a "global" equilibrium which does not imply that the structure is not locally collapsing. 
More precisely we propose that the large scales are supported by a combination of rotation and turbulence which sets the $M$-$R$ relation.

Many numerical studies of star or core formation inside molecular clouds have been performed to understand the impact of turbulence and gravity on star formation and the origin of the IMF.
For example, studies such as \citet{Padoan07} and \citet{Schmidt10} looked for gravitationally unstable clumps inside hydrodynamic or MHD simulations of a piece of molecular cloud to produce a mass spectrum and compared to theories.
\citet{Girichidis11, Girichidis12a} compared the clusters formed in 100 M$_\odot$ clouds with various initial density profile and types of turbulence using AMR hydrodynamical simulations. 
\citet{Klessen98,Klessen00} simulated a piece of clumpy molecular cloud with periodic boundary conditions initialized with Gaussian random field using SPH method.
They identified cores forming in clusters.
\citet{Bonnell03,Bonnell08,Smith09} performed hydrodynamical SPH simulations of $10^3$ M$_\odot$ uniform density spherical isothermal clouds and $10^4$ M$_\odot$ cylindrical clouds with a gradient in density respectively. 
They followed star formation with sink particles and focused on the hierarchical process of cluster formation from merging of subclusters.
They stressed on the important role of interactions among stars and the role of gravitational potential created by the cluster on the formation of brown dwarfs.
\citet{Ballesteros15} use SPH to simulate a $10^3$ M$_\odot$ cloud to characterize the impact of self-gravity on the high mass end slope of the IMF.
To study  the transition from cloud to stellar cluster and the link between their properties 
and in particular the later evolution of stellar clusters, 
\citet{Moeckel10, FujiiPortegies15} performed SPH simulations of a molecular cloud for one free-fall time,
convert the dense gas to stars, and evolve with N-body simulations to form clusters.
 These studies do not  explicitly addresses in great details how the gas, which is converted into clustered stars, is assembled
and whether and how the properties of the stellar clusters are inherited  from the gas phase. 

In this paper, we focus  on understanding the gas-dominated phase of early cluster formation inside  molecular clouds,
and show that the properties of the gaseous proto-cluster are determined by those of its parent cloud and are subsequently inherited by the stellar cluster.
We focus on the transition between the global collapse that takes place at larger scales and the mechanical equilibrium at smaller scales, creating the gaseous proto-cluster.
An energy equilibrium is reached for this gas-dominated object through gravo-turbulent interaction,
and its high density provides a favorable environment for star formation.
Certain characteristics of the stellar cluster are therefore determined at the gas-dominated phase. 

More precisely, 
this work aims at characterizing the gaseous proto-cluster formed in molecular clouds collapsing under self-gravity.
We perform a series of simulations of gaseous proto-cluster formation inside molecular cloud for various levels of the initial turbulent support, 
identify such gaseous proto-cluster regions, 
and analyze their properties to examine whether these structures are indeed objects in equilibrium, and if they coincide with observations of star-forming clumps. 
We also discuss the sink particles forming therein and compare with observations of embedded clusters. 
The nature of the clusters having been known, 
we could thus investigate star formation given such recipe of a cluster environment to gain knowledge on the star formation rate, the star-forming mechanisms, the initial mass function, etc.
From a numerical point of view, 
the wide range of temporal and spatial scales concerned in star formation simulations have always been computationally challenging.
The knowledge on how a gaseous proto-cluster forms in the molecular cloud could serve as a valuable tool for initializing a more realistic star formation simulation without having to include the whole cloud into the simulation box, 
and thus could help to gain in computation time or resolution.

To better understand the simulation results, 
an analytical model is introduced in the companion paper (paper II) to account for the gaseous proto-cluster formation in a collapsing molecular cloud, which reaches quasi-stationary equilibrium. 
Similar ideas following \citet{Hennebelle12} is used to derive the virial equilibrium for an accreting system.
As a natural consequence of angular momentum conservation during collapse, 
rotation is seen in observations \citep{HB12,Davies11,Mackey13} and simulations.
The system is therefore no longer spherically symmetric and we decompose the energy equilibrium of the proto-cluster into two dimensions.
This model provides a link from cloud properties to that of the cluster,
and determines the cluster mass, size, velocity dispersion, and the rotation.
The mass-size relation inferred from observations \citep{Fall10, Urquhart14} is very closely reproduced,
provided that the parent molecular cloud is initially in a state close to virial equilibrium.

The paper is organized as follows.
We first describe in \S 2 how the simulations are performed to investigate proto-cluster formation in molecular clouds. 
In \S 3, we explore different methods for identifying the cluster in the simulation results.
Both gas kinematics and sink particle distribution are considered.
The results are compared to observational measurements.
The energy equilibrium and internal properties of the proto-cluster are discussed in \S 4.
Finally we make some remarks and conclusions in \S 5.


\section{Simulations}
We perform numerical simulations of a collapsing molecular cloud under self-gravity with the adaptive mesh refinement (amr) magneto-hydrodynamics code RAMSES \citep{Teyssier02, Fromang06},
which uses a Godunov's scheme and constrained transport method to ensure null magnetic field divergence.
We first introduce the initial conditions of the cloud and the physical processes considered in this paper.
We also give numerical details of the simulations, and some qualitative results are presented at the end.

\begin{figure*}
\centering 
\begin{subfigure}{\textwidth}
\includegraphics[trim=10 110 10 120,clip,width=\textwidth]{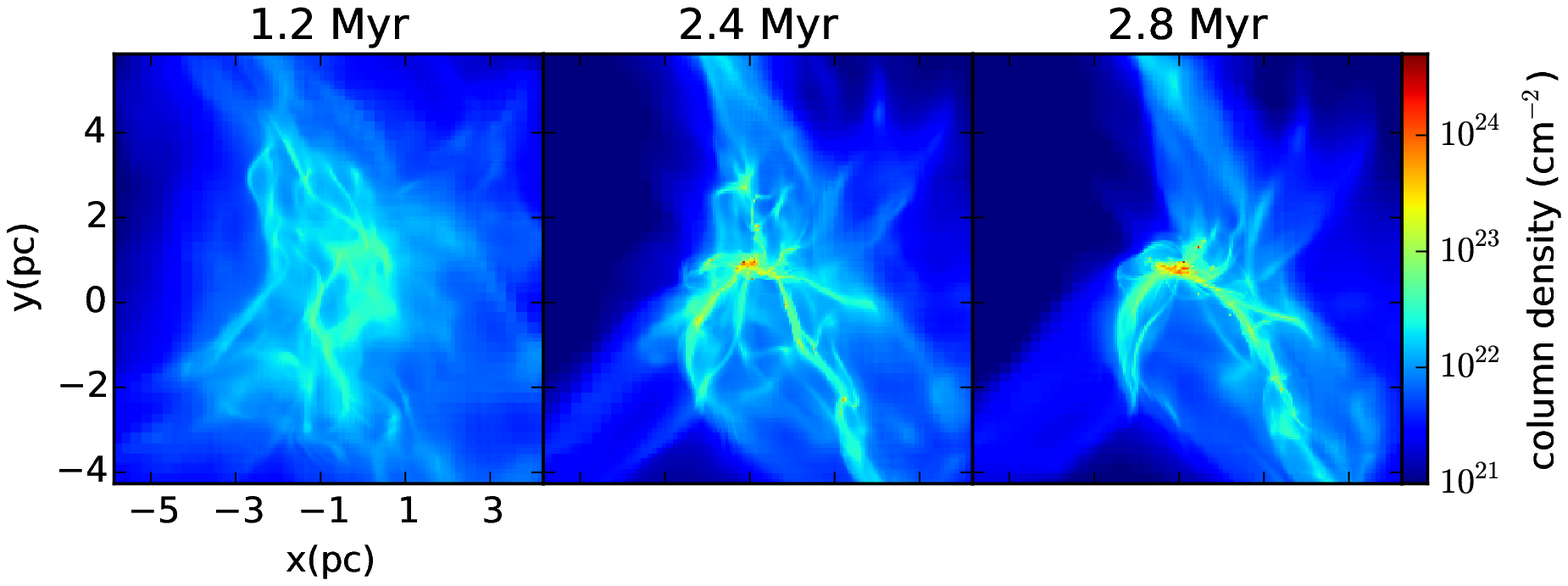}
\end{subfigure}
\begin{subfigure}{\textwidth}
\includegraphics[trim=10 110 10 120,clip,width=\textwidth]{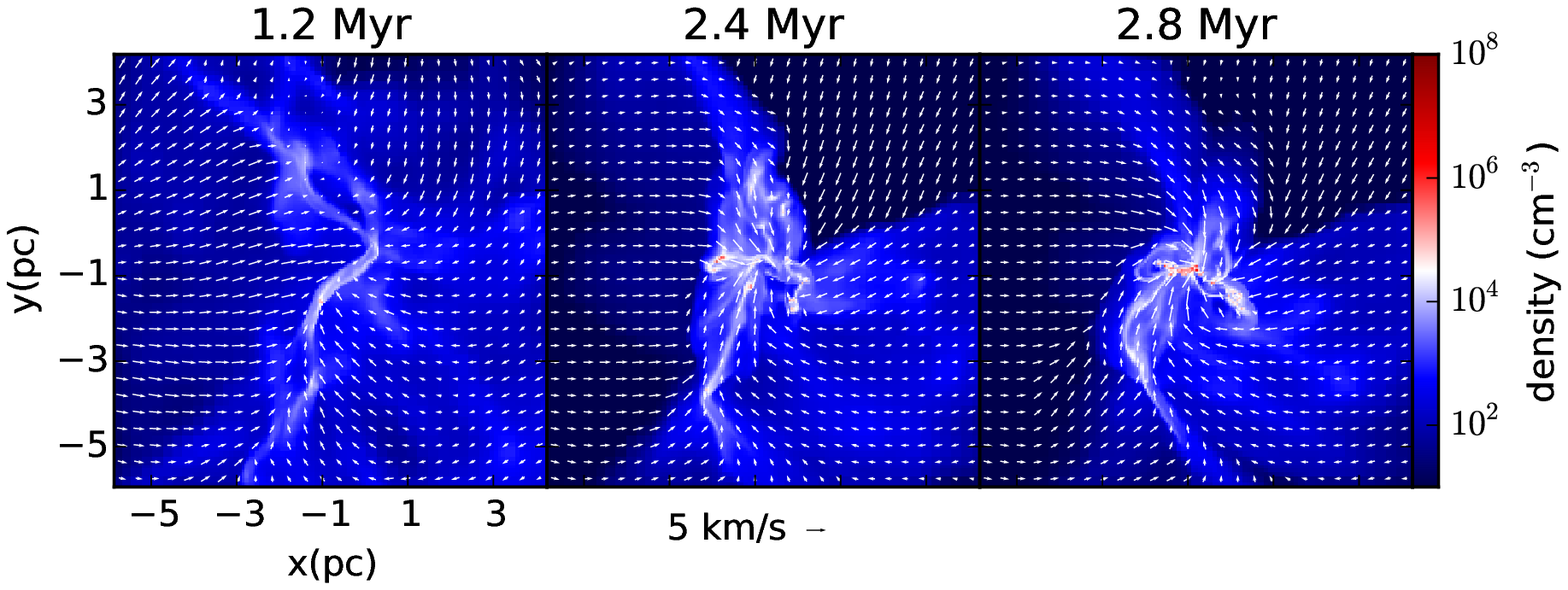}
\end{subfigure}
\caption{Zoomed images of run B at times 1.2, 2.4, 2.8 Myr,
showing the central dense objet. 
Upper panel: column density. 
Lower panel: velocity field over-plotted on density sliced map.
The origin of the coordinate corresponds to the center of the computational box,
and the figures are re-centered such that the densest cell in the rightmost panel is at the center of the image.
Sink particles are not shown.}
\label{simu_10}
\end{figure*}

\subsection{Physical processes and initial condition}
The evolution of the cloud is governed by ideal magneto-hydrodynamic equations,
with a cooling function of the ISM considered, as that described by \citet{Audit05}.

A series of molecular clouds having different levels of  initial turbulence are being performed.
Simulations are initialized with a Bonnor-Ebert-like spherical cloud of $10^4$ solar mass,
having a density profile $\rho(r) = \rho_0 /\left[ 1+\left({r \over r_0}\right)^2\right]$ with 10 times density contrast between the center and the edge,
where $\rho_0 = 822 ~\mathrm{H} ~\mathrm{cc}^{-1}$ and $r_0 = 2.5 ~\mathrm{pc}$.
The cloud has 15 pc diameter and the computational box is twice the size of the cloud, 
i.e. 30 pc in each dimension.
The space surrounding the cloud is patched with diffuse gas of uniform density $\rho_0/100$.
The initial temperature is set by the cooling function and  it is about 10 K in the dense gas.
A turbulent velocity field is generated with a kolmogorov spectrum with random phase and is scaled in such a way 
that the initial 
Mach numbers are 2.7, 6, 7.3,  and 10, respectively. 
A weak magnetic field with uniform mass to flux ratio in the $x$-direction is applied.
Its  largest value is  about $8 ~\mathrm{\mu G}$ at the cloud center and 
reaches about $3 ~\mathrm{\mu G}$ at the cloud edge.

The ratios of characteristic time scales at the central plateau region ($r<r_0$) are used as parameters to specify the initial conditions.
In all runs, we use  the ratio of free-fall time to sound crossing time $t_\mathrm{ff}/t_\mathrm{sct} = 0.15$, 
and the ratio of free-fall time to Alfv\'en crossing time $t_\mathrm{ff}/t_\mathrm{act} = 0.2$, which corresponds to a  value of the mass-to-flux over critical
mass-to-flux ratio of about 8. 
This implies that both thermal and magnetic energy are small compared to the gravitational potential energy.
The ratio between free-fall time and turbulent crossing time $t_\mathrm{ff}/t_\mathrm{vct}$ is varied in each run,
giving different levels of kinetic support against self-gravity.
In Table \ref{table_params}, we give the values of $t_\mathrm{ff}/t_\mathrm{vct}$ 
as well as the corresponding turbulent rms velocity, the Mach number, and the virial parameter $\alpha_\mathrm{vir} = {2E_\mathrm{kin}/E_\mathrm{grav}}$ to characterize the initial energy state of the molecular cloud.

\begin{table}
\caption{Simulation parameters: The cloud is specified with varying level of the ratio $t_\mathrm{ff}/t_\mathrm{vct}$.
In the table we list corresponding turbulent velocity and the virial parameter $\alpha_\mathrm{vir} = {2E_\mathrm{kin}/E_\mathrm{grav}}$ for each run.}
\label{table_params}
\centering
\begin{tabular}{c l c c c c }
\hline\hline
Label   & $t_\mathrm{ff}/t_\mathrm{vct}$  &  \makecell{$v_\mathrm{rms}$ \\(km/s)} & Mach  & $\alpha_\mathrm{vir}$ \vspace{.5mm}\\
\hline
A  &  0.4  &  0.6 & 2.7  & 0.12\\
B  & 0.9   & 1.5& 6.0 &  0.64 \\
C  & 1.1  & 1.8  & 7.3& 0.96\\
D  & 1.5 & 2.4 & 10&  1.78\\
\hline
\end{tabular}
\end{table}

\begin{figure*}
\centering 
\begin{subfigure}{\textwidth}
\includegraphics[trim=10 110 10 120,clip,width=\textwidth]{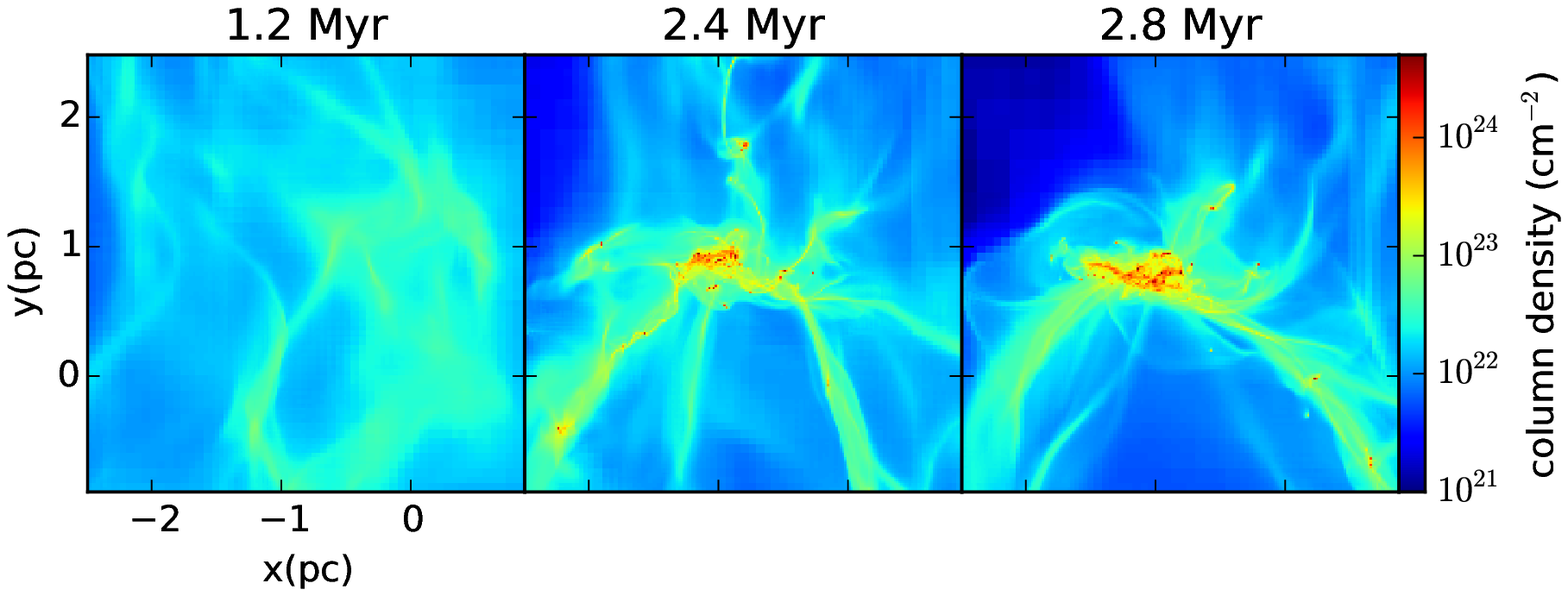}
\end{subfigure}
\begin{subfigure}{\textwidth}
\includegraphics[trim=10 110 10 120,clip,width=\textwidth]{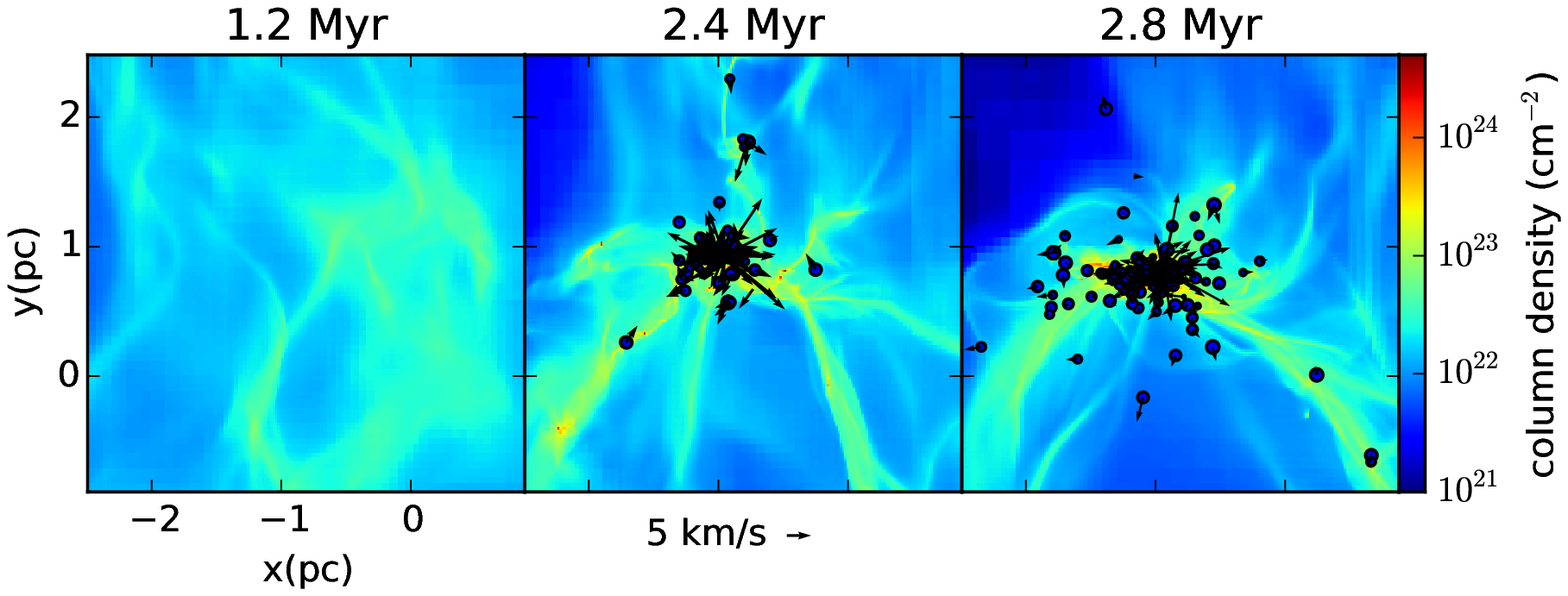}
\end{subfigure}
\caption{Zoomed images of run B at times 1.2, 2.4, 2.8 Myr. 
Upper panel: column density. 
Lower panel: sink particles over-plotted on column density. 
The circles represent the sink particles with the size proportional to the log of their masses,
and the arrows indicate their velocity.
}
\label{simu_3_col}
\end{figure*}

\subsection{Numerical setup}
We start with a $128^3$ base grid and allow $7$ levels of refinement (equivalent to $16384^3$) using amr scheme to ensure a resolution of 10 grids per jeans length.
When increasing by one level, a cell is subdivided into 2 in each dimension.
This gives a resolution of 0.23 pc on the coarse grids and refinement down to 0.002 pc (or 400 AU) at the densest regions.
Neumann boundary conditions with imposed zero gradients are used, which allows  the gas to outflow from the computational box.

Sink particles are used in our simulations as described by \citet{Bleuler14},
the densest region unresolved with a fluid description is replaced with a sink particle to economize computational power and to follow accretion onto Lagrangian mass.
They are formed when density exceeds $10^8 ~\mathrm{H} ~\mathrm{cc}^{-1}$,
while the flow is convergent and super-virial. The sink radius is equal to 4 computational cells at the finest resolution. 
After their formation, the sink particles are capable of accreting mass from their surrounding.
As these high density regions represent possible star formation sites,
they furnish dynamical and statistical hints on stellar cluster properties.
We caution that although our resolution is very small with respect to the cluster size, insuring good description of the cluster 
itself, the sink particle size is too large for the sinks to represent individual stars. The typical mass 
of our sink particles is on the order of $\simeq 10$ M$_\odot$.

\subsection{Qualitative presentations}
\begin{figure*}
\centering 
\begin{subfigure}{\textwidth}
\includegraphics[trim=10 110 10 120,clip,width=\textwidth]{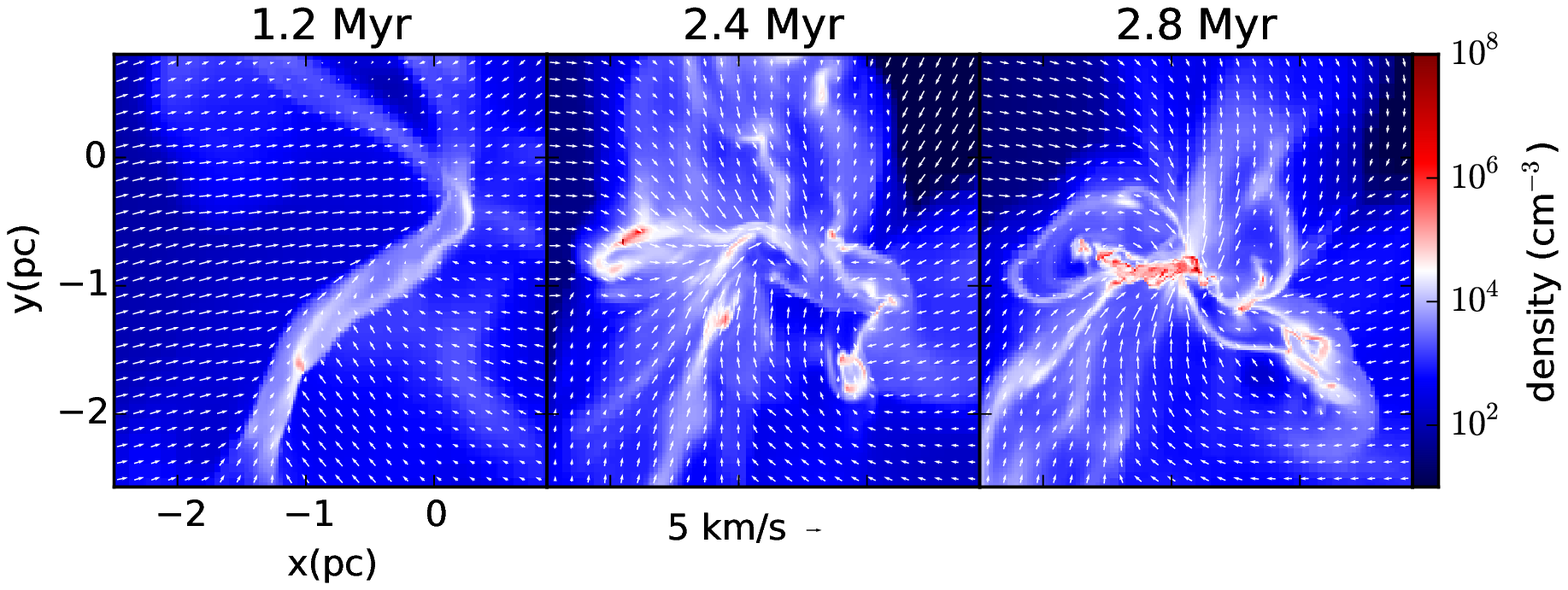}
\end{subfigure}
\begin{subfigure}{\textwidth}
\includegraphics[trim=10 110 10 120,clip,width=\textwidth]{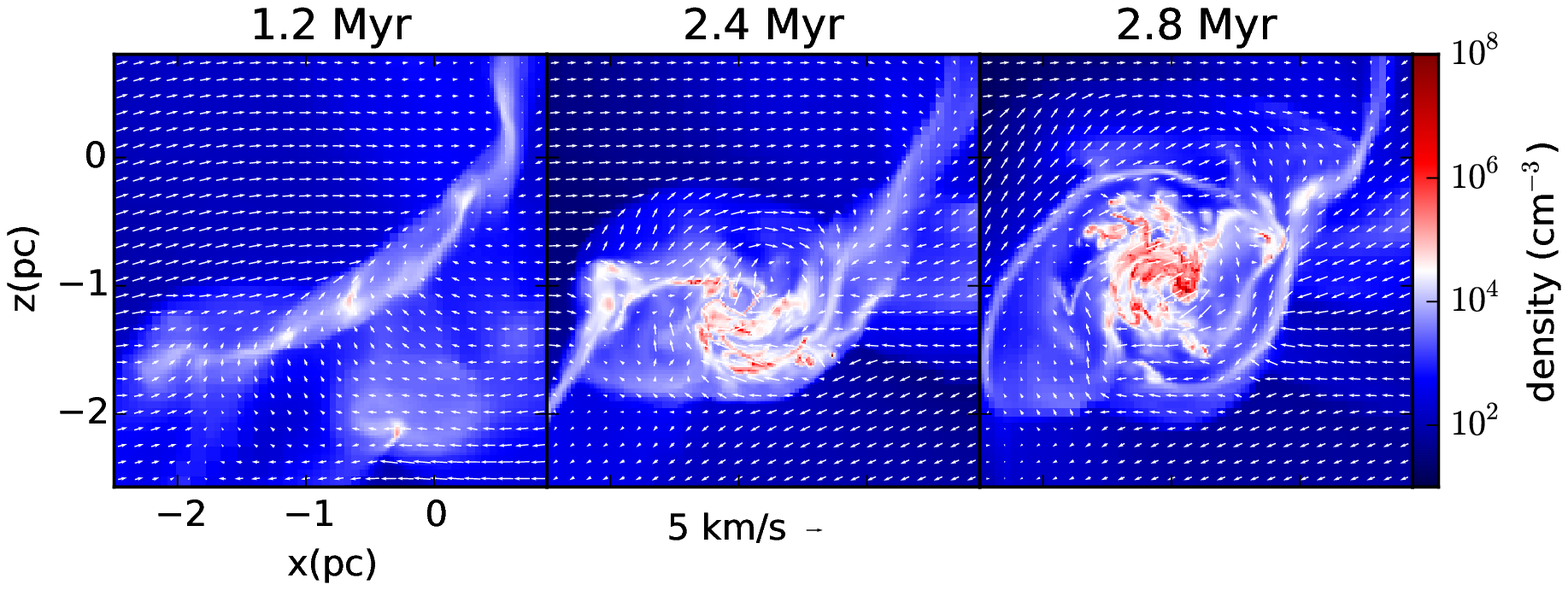}
\end{subfigure}
\caption{Zoomed images of run B at times 1.2, 2.4, 2.8 Myr focusing on the central dense objet,
with two views showing the flattened shape. 
Upper panel: velocity field over-plotted on density sliced map, edge-on view.
Lower panel: velocity field over-plotted on density sliced map, face-on view of the flattened rotating proto-cluster.
}
\label{simu_3_den}
\end{figure*}

Column density maps and velocity field over-plotted on density slice maps are shown in Fig. \ref{simu_10} for run B.
Figures \ref{simu_3_col} and \ref{simu_3_den} show a zoomed view of the central region where the proto-cluster is forming.
The cloud is globally collapsing,
while filamentary structures are forming under gravo-turbulent interactions.
When zooming in towards the central region,
we see a parsec-scale prominent entity of relatively high density emerging, whose size is slowly varying in time.
A general rotational motion becomes evident as the collapse proceeds, 
since there is no dissipative mechanisms efficient enough to remove large amount of angular momentum at this scale.
The global infalling motion is noticeably reduced upon striking the central objet,
forming highly irregular shocks at the seemingly ill-defined border.

To have a fully physical description of a realistic self-regulation of star mass accretion and cluster formation,
stellar feedbacks such as radiations \citep{Bate09b, Price09, commercon11, Bate12, Krumholz12}, 
protostellar jets and outflows \citep{Wang10, Nakamura11, Dale13c, Nakamura14, Federrath14, Dale15}, 
HII regions \citep{Krumholz07, Peters10, Dale13b, Dale13a, Geen15, Dale15}, 
and supernovae \citep{Iffrig15, Walch15} should be incorporated.
In this work we limit ourselves to the simple scenario without feedback mechanisms,
in which nothing except the depletion of cloud gas can stop sink accretion.
We therefore stop the simulations when about half of the initial cloud mass has been accreted onto the sink particles,
since the results are less likely to be physical among this stage.
This corresponds to physical time of about 3 million years.
While some of the sink particles form in the filamentary structures feeding the central cluster,
 most of the sinks are observed to form within the gaseous proto-cluster,
 hence strongly suggesting that this central region is an important site for understanding the star properties such as 
the IMF.


\section{Inferring the proto-cluster radius from simulation results}
\label{simu}
As previously depicted, although a prominent structure forming at the center of the cloud is exhibited, 
it is indeed highly irregular.
With a view to attain a global description of the proto-cluster, 
we present our strategies to infer the cluster radius in this section.
Two approaches are exploited and compared: 
the radius of the proto-cluster is determined with the gas kinematics and the sink particle distribution respectively.

\subsection{The gaseous proto-cluster}
In the density maps shown in Fig. \ref{simu_3_den}, 
we see a relatively high density flattened structure forming at the center, dominated by rotation.
The infalling motion is stopped by a shock at its highly irregular border, where we see abrupt change in flow directions.
This structure is somewhat stationary with a constant size, while mass is accreting.
Using the gas properties, i.e. the density, the velocity, the turbulence, the rotation, etc,
we search a way to identify this region in order to study its global characteristics.
The turbulent nature prevents us from studying the detailed gas behavior in a simple way.
One possibility to get rid of pronounced fluctuations is to calculate integrated quantities in a given region,
and thus characterizing the proto-cluster.
We therefore proceed in two steps.
We first determine a series of ellipsoids, and then decide which one best represents the cluster.

\subsubsection{Step 1: integrated gas properties of an ellipsoid}
In order to define the cluster,
we calculate the center of mass, the total angular momentum, and the moment of inertia
in an oblate ellipsoidal region of semi-major axis $R$ and semi-minor axis $H$.
For a series of $R$ values,
we compute iteratively to find the corresponding $H$ by replacing
\emph{1)} the geometrical center by the center of mass, 
\emph{2)} the minor axis by the axis of rotation,
and \emph{3)} the geometrical aspect ratio by that obtained from mass distribution, the inertia momenta,
until convergence is reached.
We find that the procedure typically converges in around 10 iterations leading to a variation of less than $10^{-4}$ and gives reasonable results.
The ellipsoids attained therefore satisfy the following criteria:
\begin{subequations}
\begin{align}
\vec{x}_\mathrm{center} &= \frac{\sum \vec{x}_i m_i}{\sum m_i} \\
\vec{a}_H &\sslash \sum m_i \vec{v}_i \times \vec{r}_i \\
({H \over R})^2 &= {\lambda_1 \over \sqrt{\lambda_2 \lambda_3}}
\label{converg} 
\end{align}
\end{subequations}
The quantities $x$ and $m$ with subscript $i$ indicate the position and mass of each cell inside the ellipsoid,
and $\vec{a}_H$ is the vector representing the direction of the minor axis.
The eigenvalues of the moment of inertia are $\lambda_{1,2,3}$, in increasing order.
This procedure defines a series of self-consistent ellipsoids contained in each other.
The disordered flow nature makes that the ellipsoids are not necessarily similar nor aligned.

\subsubsection{Step 2: the fitting procedure}
A cloud forming a central cluster features a collapsing envelope and a quasi-stationary core with minor infalling motion.
To distinguish between the two, we define a quantity
\begin{eqnarray}
W_0 = \int_{V(R)} \vec{v}\cdot \vec{n} ~dm ~/~ \|\int_{V(R)} \vec{v}\times \vec{n}~dm\|
\end{eqnarray}
for the series of ellipsoids of different size to observe the change in the collapsing motion,
where $\vec{v}$ is the velocity and $\vec{n}$ the unit vector pointing from the ellipsoid center to the cell position.
While the envelope is expected to be globally infalling,
the gaseous proto-cluster itself should be relatively stationary with more prominent rotation. 
The quantity $W_0$ should therefore follow different radial dependence inside and outside  the proto-cluster. 
We show by simple arguments that its numerator $F_0 = \int \vec{v}\cdot \vec{n} ~dm $ follows different radial dependences inside and outside the gaseous proto-cluster: 
by making analogy to the collapsing phase of the Larson-Penston solution of a sphere \citep{Larson69, Penston69},
where the mass flux is conserved for all radii ($\rho \propto r^{-2}$ and $v_\mathrm{inf} \propto constant$) in a stationary state,
we conclude that $F_0$ should be roughly proportional to $R$ in the envelope.
\begin{eqnarray}
\rho v_\mathrm{inf} 4\pi r^2 &\propto& r^0 \nonumber\\
F_0 = \int \rho v_\mathrm{inf}  dV &\propto& R 
\end{eqnarray}
As for the region inside the gaseous proto-cluster,
the assumption of uniform density implies that the density increasing rate is also uniform.
We deduce that $F_0 \propto R^4$.
\begin{eqnarray}
d_r (\rho v_\mathrm{inf} 4\pi r^2)  =  \dot{\rho} 4\pi r^2 &\propto& r^2 \; (\dot{\rho} \text{ is a constant of} \;r)  \nonumber\\
F_0 = \int \rho v_\mathrm{inf} dV &\propto& R^4 
\end{eqnarray}

\begin{figure}[]
\centering
\includegraphics[width=0.5\textwidth]{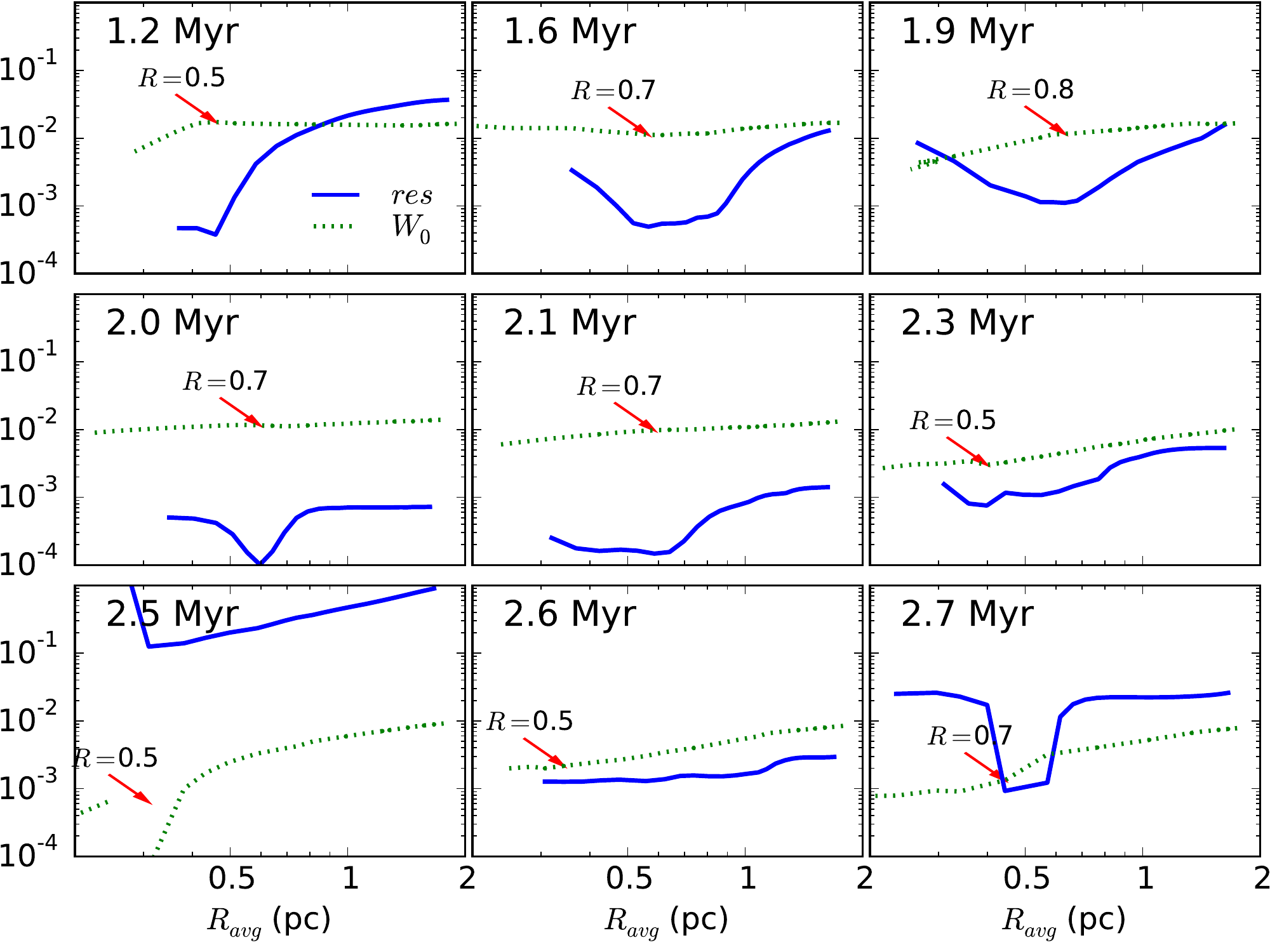}
\caption{Examples of the piecewise fit at nine time steps for the run with $t_\mathrm{ff}/t_\mathrm{vct}$=0.9. 
The normalized residual is plotted as function of cluster characteristic radius $R_\mathrm{avg} = (R^2H)^{1\over 3}$ in blue solid curves, of which the optimal value is such that the fit is the best.
The infall moment $W_0(R_\mathrm{avg})$ is shown in arbitrary units with green dotted curves. 
The gaseous proto-cluster radius at which there exists a change in slope is indicated with a red arrow and the size of the semi-major axis is marked.
}
\label{fits}
\end{figure}

Though these are not exact derivations for a configuration with rotation, 
a difference in radial dependence is clear when a quasi-stationary core exists. 
We therefore propose a simple piece-wise power-law description which differentiates the flow properties inside and outside the gaseous proto-cluster.
As shown in Fig. \ref{fits},
the infall moment exhibits a change in slope in log-scale.
Inside the cluster, 
the gas is no longer necessarily collapsing and $W_0$ can even change signs.
A piece-wise fit is performed and the residual calculated to define the semi-major axis $R_\ast$ of the gaseous proto-cluster. 
\begin{subequations}
\begin{align}
W_{0,\mathrm{fit}}(R_\mathrm{avg}, r_\ast)  &= \left\{ \begin{array}{lcl} a_c R_\mathrm{avg}^{b_c}  & \mbox{for} & R_\mathrm{avg}<r_\ast \\
a_e R_\mathrm{avg}^{b_e}  & \mbox{for} & R_\mathrm{avg}>r_\ast \end{array} \right. \\
res(r_\ast) &={1\over n_R} \sum_R \left[ \frac{W_{0,\mathrm{fit}}(R_\mathrm{avg}, r_\ast) - W_0(R)}{W_0(R)} \right]^2    \\
R_\ast &= \arg \min{res(r_\ast)} 
\end{align}
\end{subequations}
The average radius of the ellipsoid $R_\mathrm{avg} = (R^2H)^{1\over 3}$ is used to perform the power-law fit since it is the most proper value to represent the ellipsoid. 
The choice of this quantity is discussed in appendix \ref{Ravg}.
The inner part of the fitted function is put to zero when there are fluctuations around zero, 
and the negative points are omitted when fitting the outer part.
The gaseous proto-cluster size is defined as the radius at which the minimal local minimum occurs.
This normalized residual of the best fit is usually smaller than $10^{-3}$,
and its square root implies a few percent error of the fit.
The fit residual is large when there exist changes in sign of $W_0$ due to our definition of the functions,
but the radius is on the contrary very well defined in such cases.
A local minimum of the fit residual cannot always be found.
This might be due to the fact that the gas is highly turbulent 
and that there exists a fluctuating shock zone between the envelope and the gaseous proto-cluster, 
which deteriorates the fit if it is included into either of the two domains.
Nonetheless, when it exists, it marks well the transition between the two regimes. Moreover as described below, 
the results obtained are entirely reasonable
with the gaseous proto-cluster radius inferred  indicating generally well the dense central region that we see.

\begin{figure*}[t!]
\centering
\begin{subfigure}{.35\textwidth}
\includegraphics[width=\textwidth]{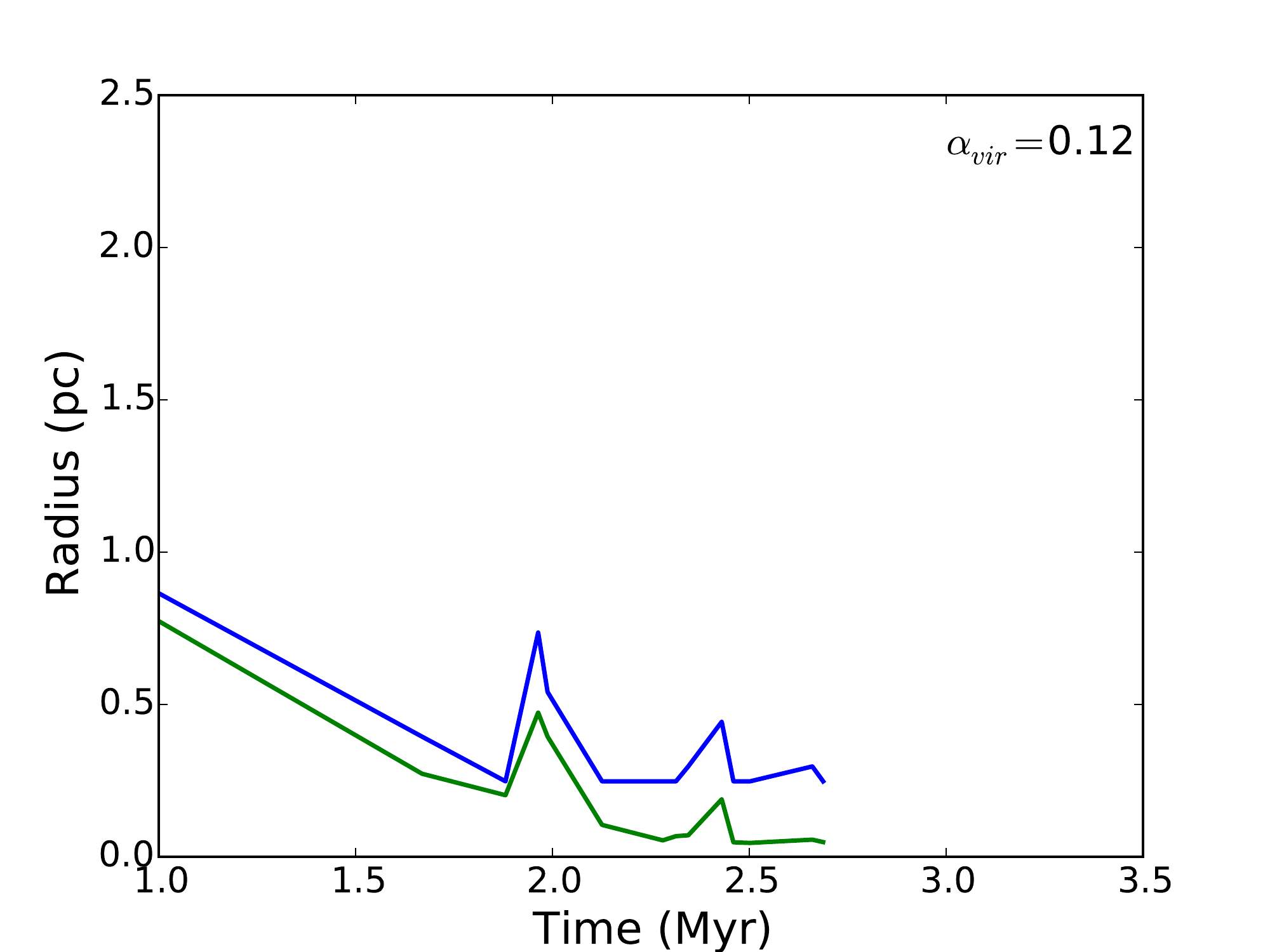}
\end{subfigure}
\begin{subfigure}{.35\textwidth}
\includegraphics[width=\textwidth]{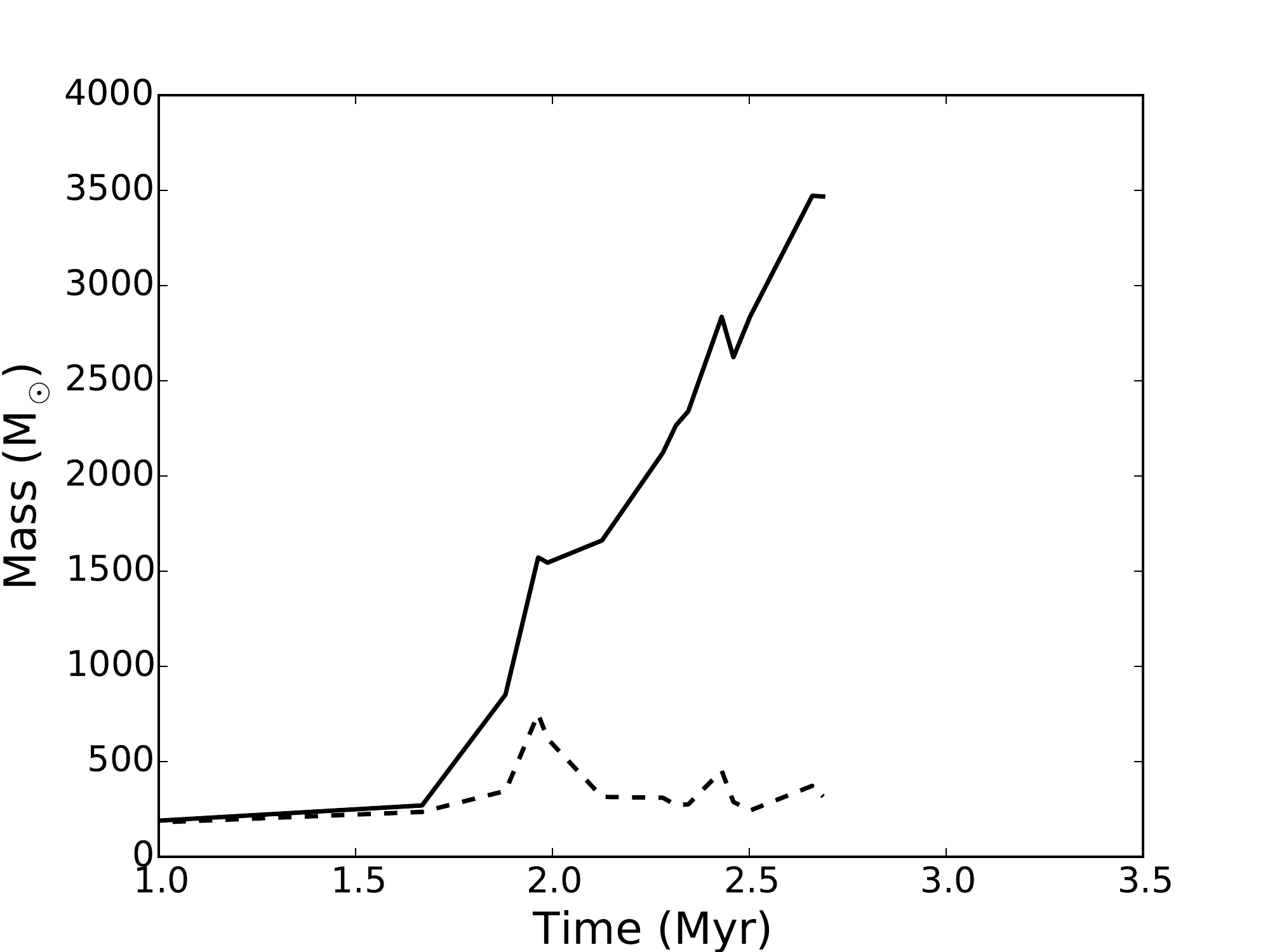}
\end{subfigure}
\begin{subfigure}{.35\textwidth}
\includegraphics[width=\textwidth]{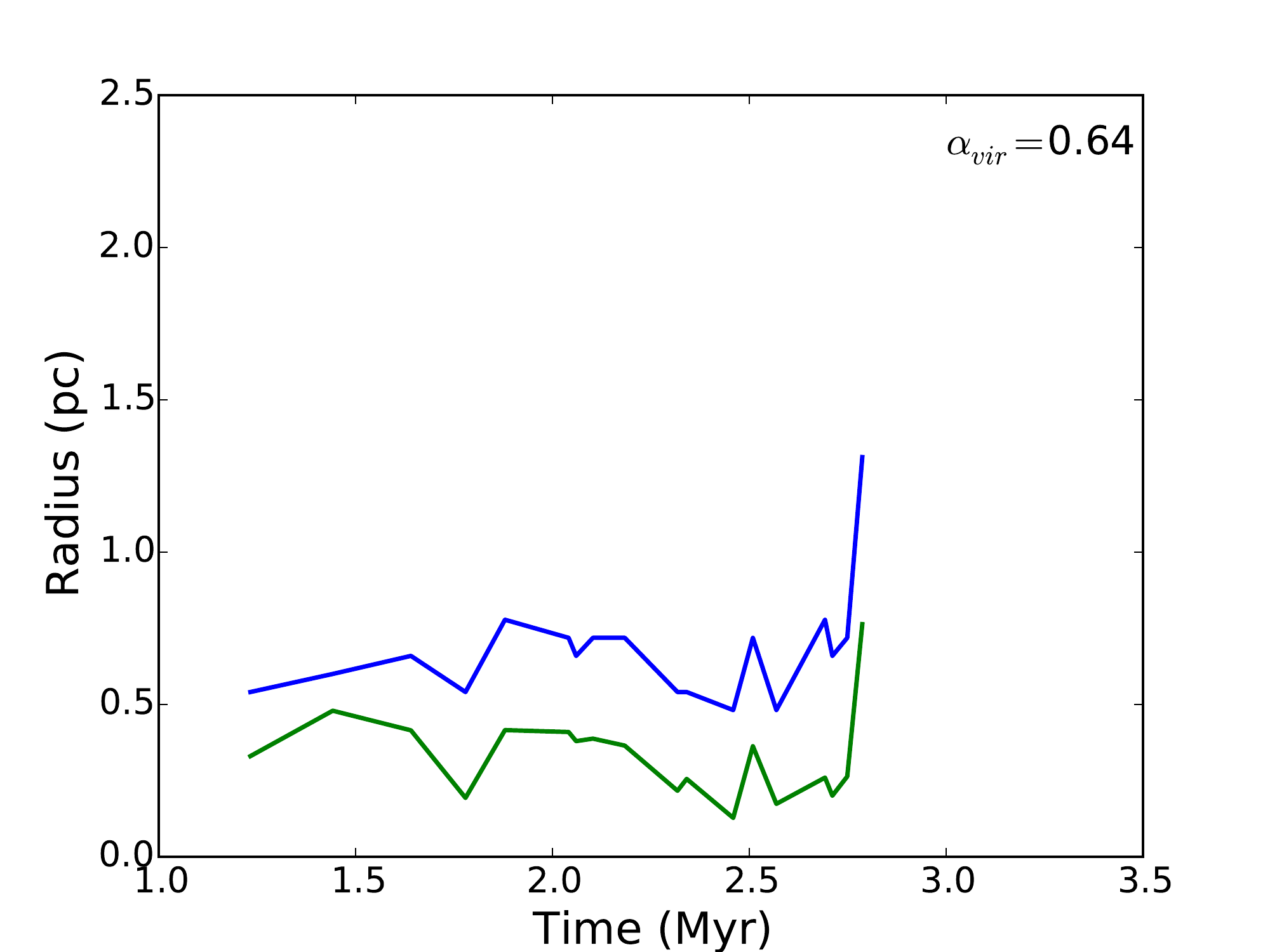}
\end{subfigure}
\begin{subfigure}{.35\textwidth}
\includegraphics[width=\textwidth]{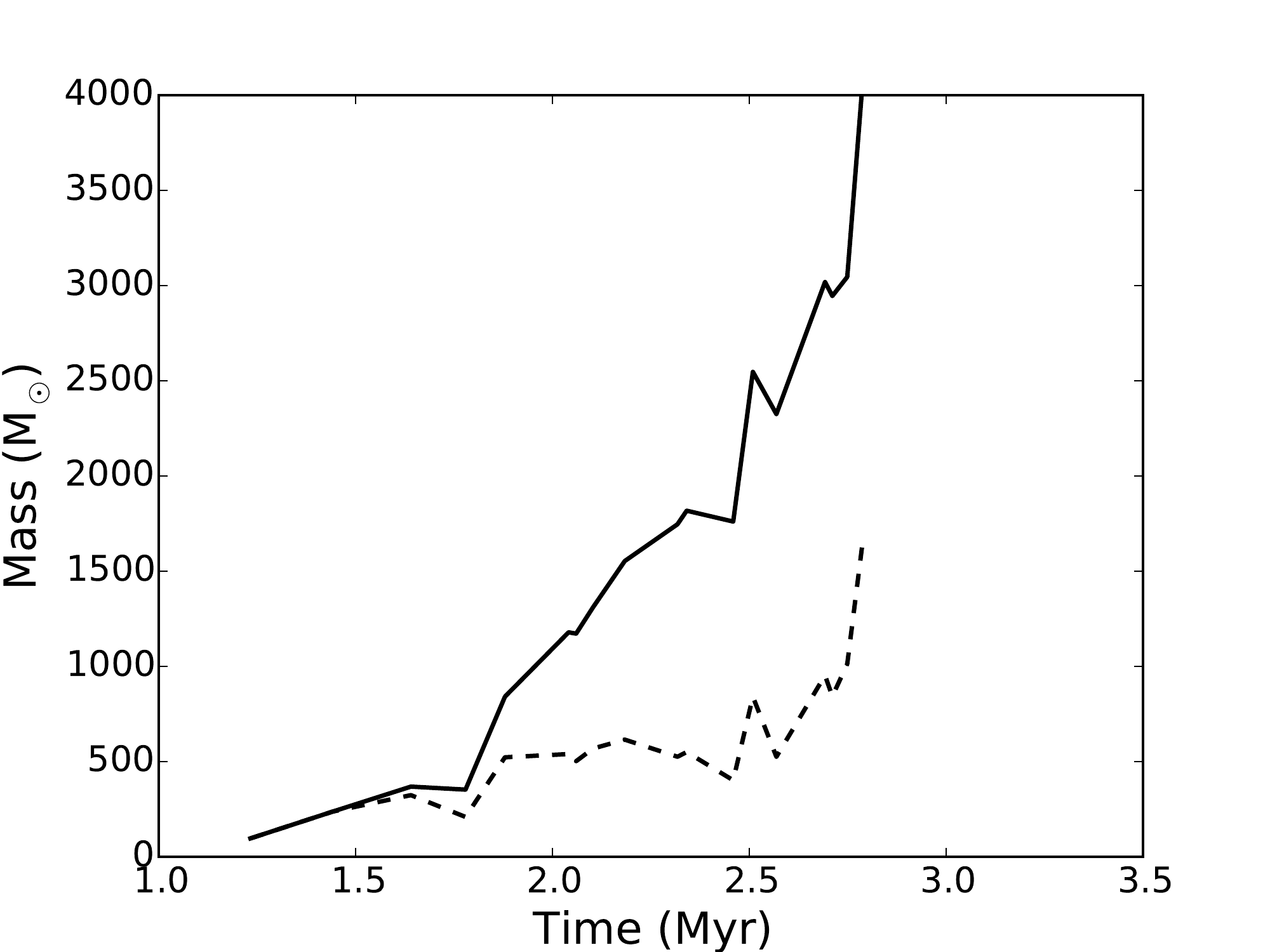}
\end{subfigure}
\begin{subfigure}{.35\textwidth}
\includegraphics[width=\textwidth]{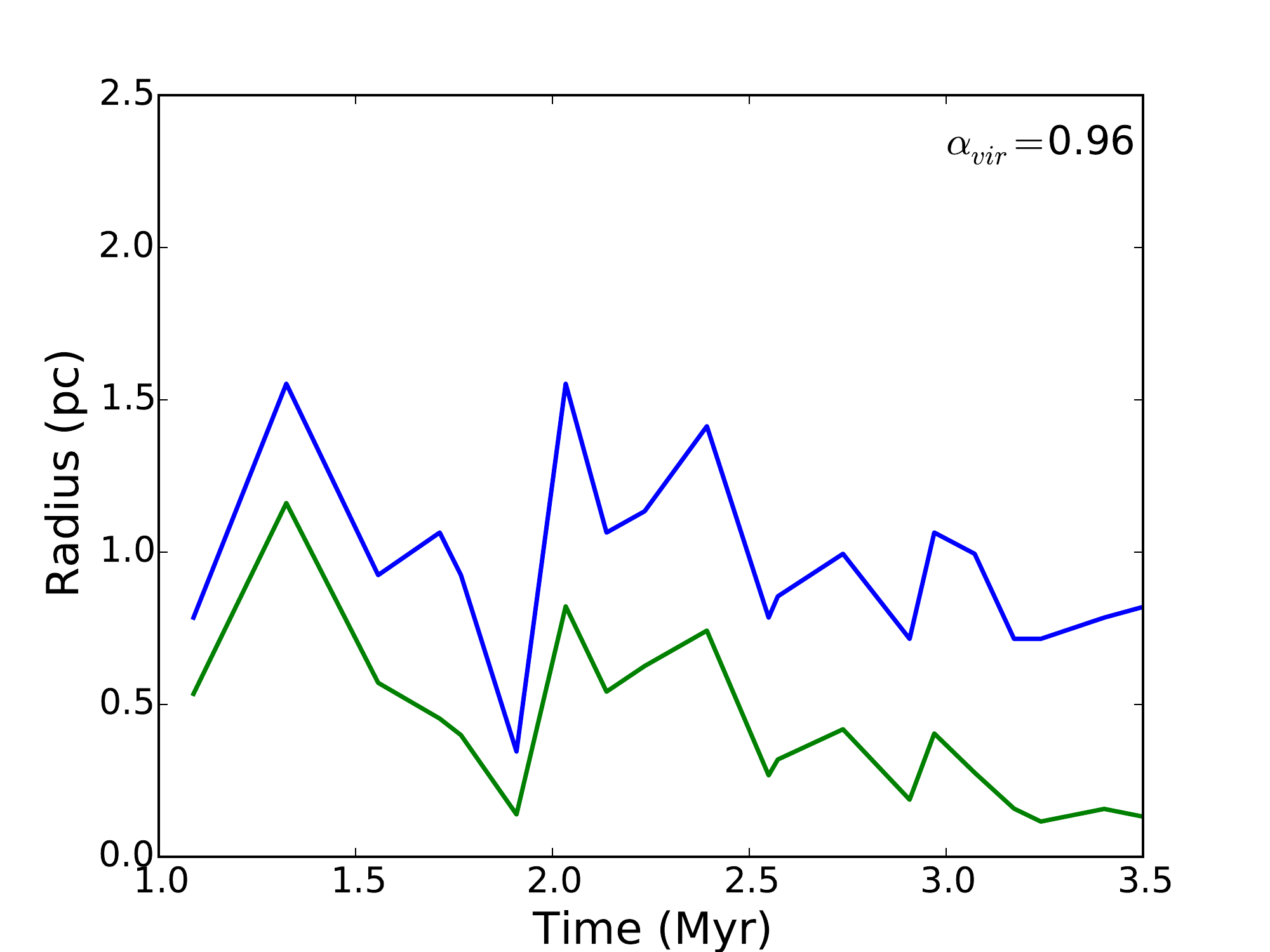}
\end{subfigure}
\begin{subfigure}{.35\textwidth}
\includegraphics[width=\textwidth]{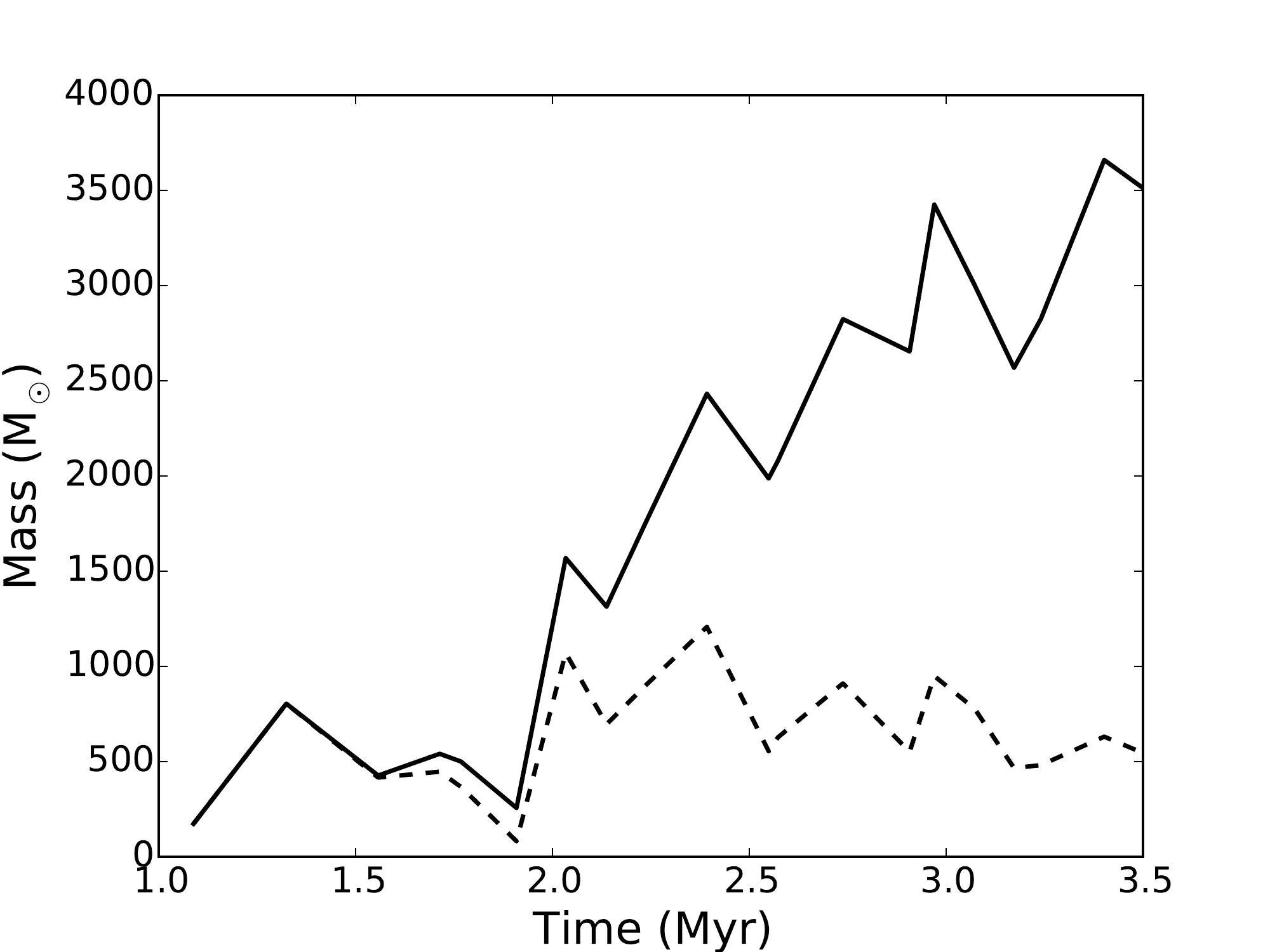}
\end{subfigure}
\begin{subfigure}{.35\textwidth}
\includegraphics[width=\textwidth]{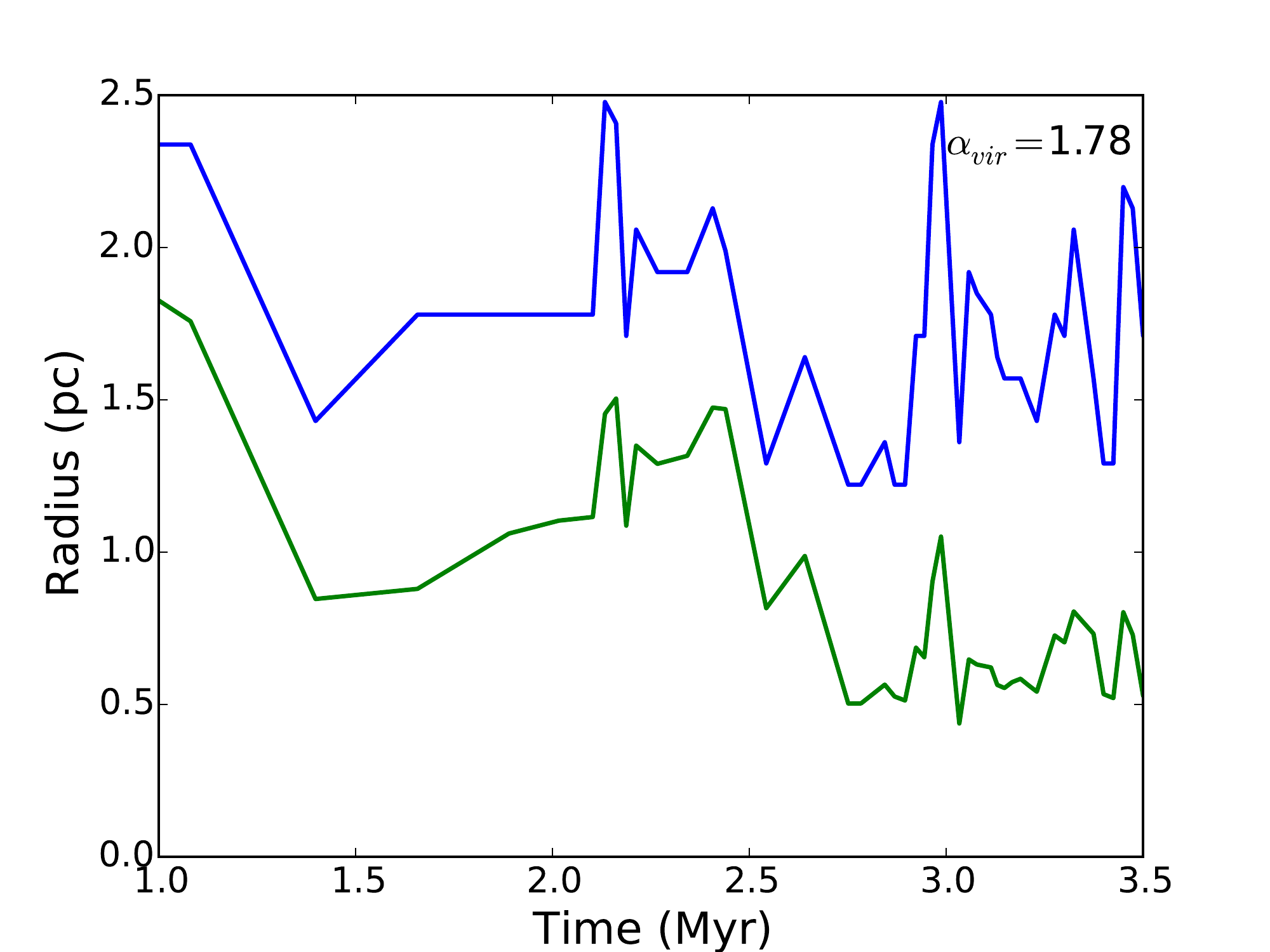}
\end{subfigure}
\begin{subfigure}{.35\textwidth}
\includegraphics[width=\textwidth]{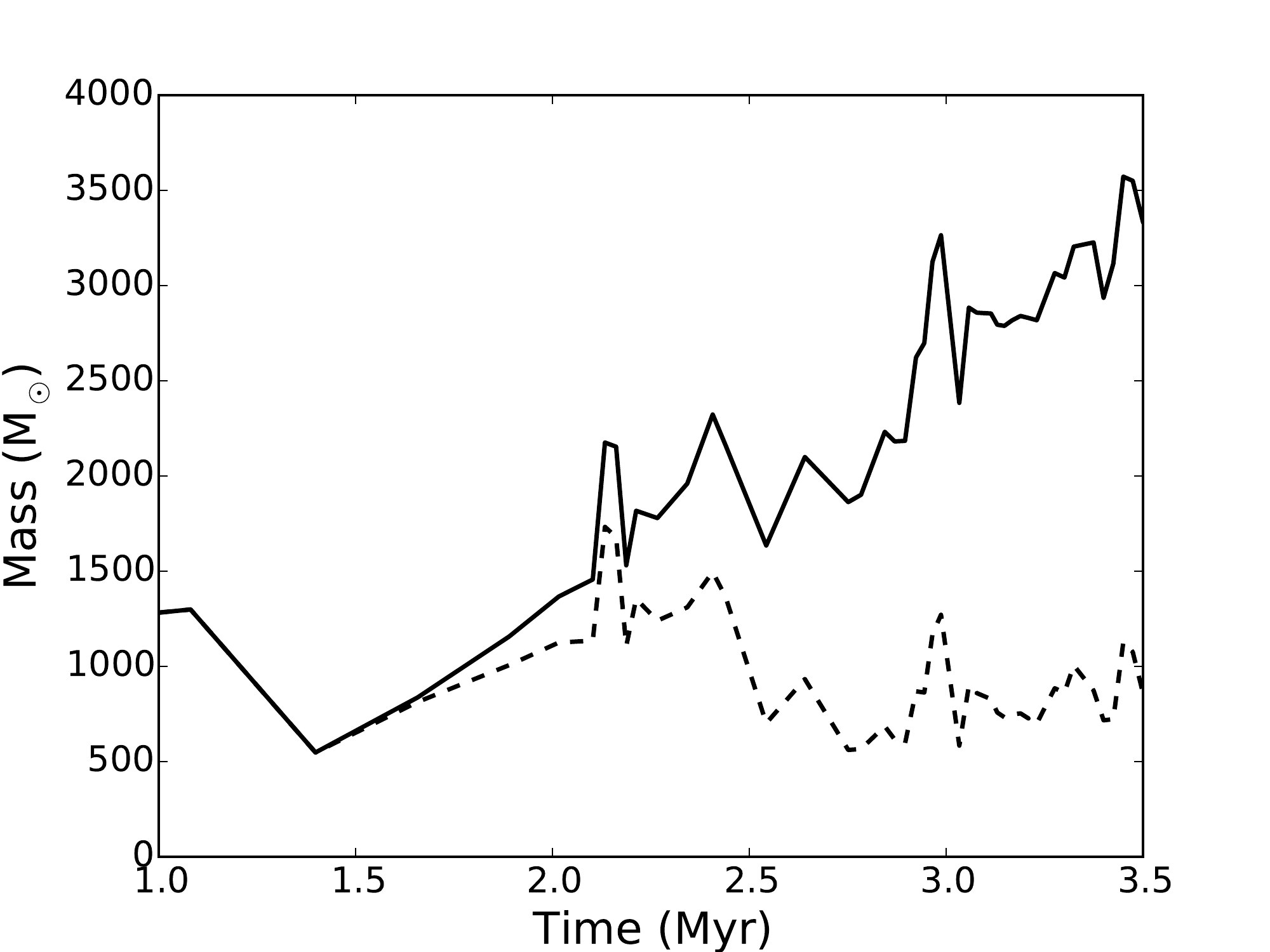}
\end{subfigure}
\caption{The time evolution of the semi-axes $R$ and $H$ of the ellipsoidal gas cluster on the left column.
On the right column, the cluster mass is shown as function of time. The two curves represent the gas mass (dashed) and that plus the sink mass (solid) inside the ellipsoidal region defined with gas kinematics.  
From top to bottom are runs A, B, C, and D with increasing levels of turbulence.
The proto-cluster size increases with turbulence level.
Most of the mass is accreted onto the sinks while the cluster mass increases in time, 
therefore the gas mass stays roughly constant. 
}
\label{rt_mt_gas}
\end{figure*}

\subsubsection{results}
The inferred evolution of size and mass of the cluster are presented in Fig. \ref{rt_mt_gas}.  
Despite its turbulent nature,
the gaseous proto-cluster size stays relatively constant while mass is accreted. 
The more turbulent the initial cloud is, 
the larger the proto-cluster is.
The mass evolution is plotted with the gas mass and total mass (sinks inside the ellipsoid included). 
The gas mass stays roughly constant while the total mass increases,
implying that most of the mass accreted onto the cluster ends up in the sinks.
As expected, the radius drops with $\alpha_\mathrm{vir}$ because of the weaker rotational and turbulent support. 
This is reproduced by the analytical model developed in paper II.
It is important to note that these simulations are done without feedback from the sink particles as a result of core or star formation.
With feedback considered, the sink formation and accretion could be substantially delayed and/or reduced.

\subsection{The embedded sink cluster}
\subsubsection{A simple method}
As recalled previously, in the simulations, sink particles are formed to follow regions of concentrated mass.
Those regions are the places where stars are likely to form,
therefore the sink particle distribution is representative of the stellar cluster and
 the cluster size can be inferred from the distribution of the sink particles.
This is done by computing the three eigenvalues of the rotational inertia matrix of sink particles with respect to their center of gravity.
As the eigenvalues represent the widths in three orthogonal directions, 
they reveal not only the size but also the shape of the sink particle cluster.
The rotational inertia matrix is
\begin{eqnarray}
I_\text{rot} = \sum_i m_i\begin{bmatrix}
x_i^2 & x_iy_i & x_iz_i  \\
x_iy_i & y_i^2 & y_iz_i  \\
x_iz_i & y_iz_i & z_i^2  \end{bmatrix},
\label{Irot} 
\end{eqnarray}
where $i$ is the index of sink particles. Its three eigenvalues $\lambda_1, \lambda_2, \lambda_3$ give the cluster size in three orthogonal directions:
\begin{eqnarray}
r_i = \beta \sqrt{{5 \lambda_i \over M}} , \; i = 1, 2, 3,
\label{rsink} 
\end{eqnarray}
where $M$ is the total mass of sinks.
The factor $5$ comes from the assumption that the sink mass is uniformly distributed in space, 
and a correctional factor $\beta \geq 1$ accounts for the fact that the mass distribution might not be uniform but rather centrally concentrated.

\begin{figure*}[t!]
\centering
\begin{subfigure}{.35\textwidth}
\includegraphics[width=\textwidth]{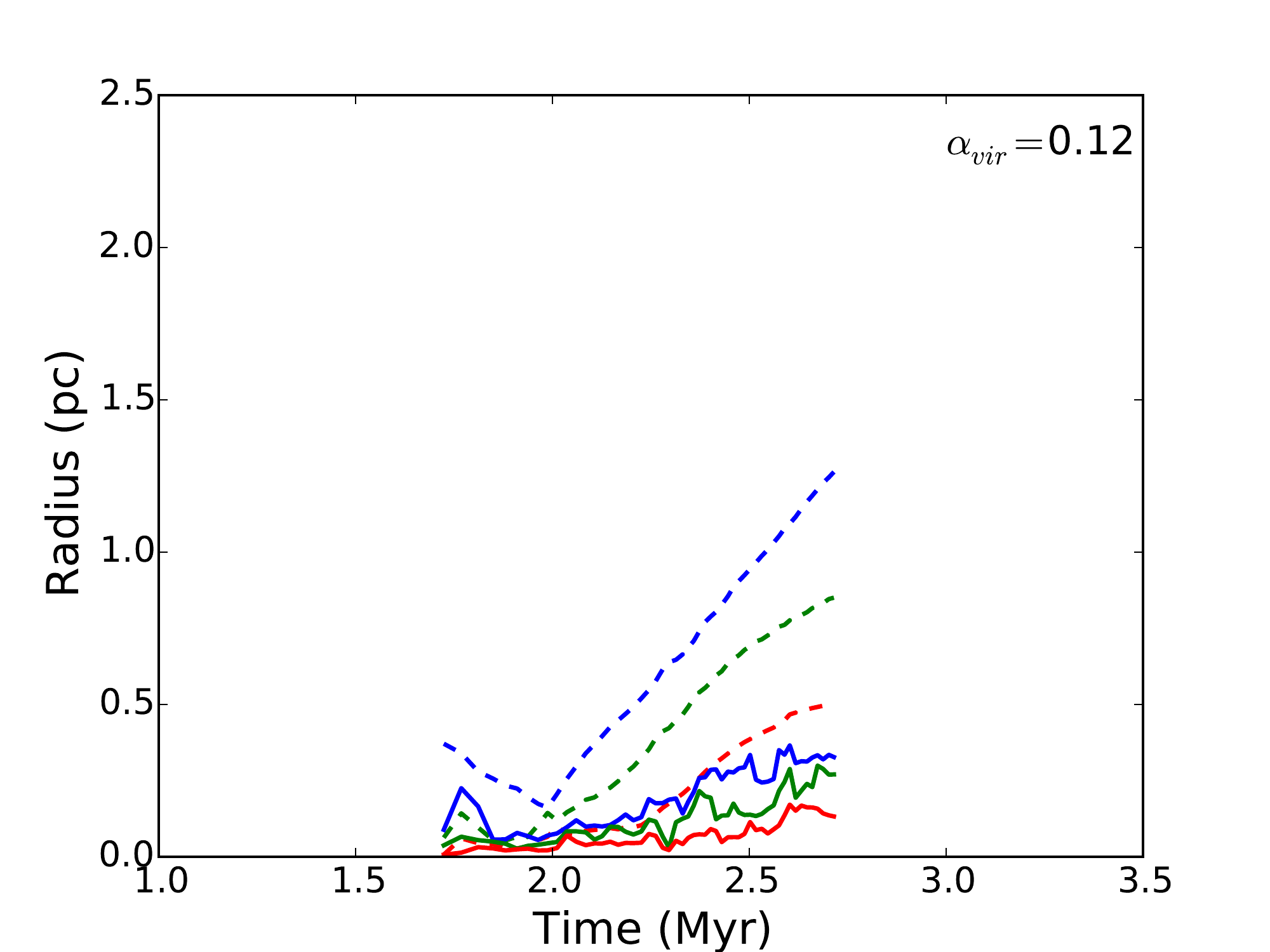}
\end{subfigure}
\begin{subfigure}{.35\textwidth}
\includegraphics[width=\textwidth]{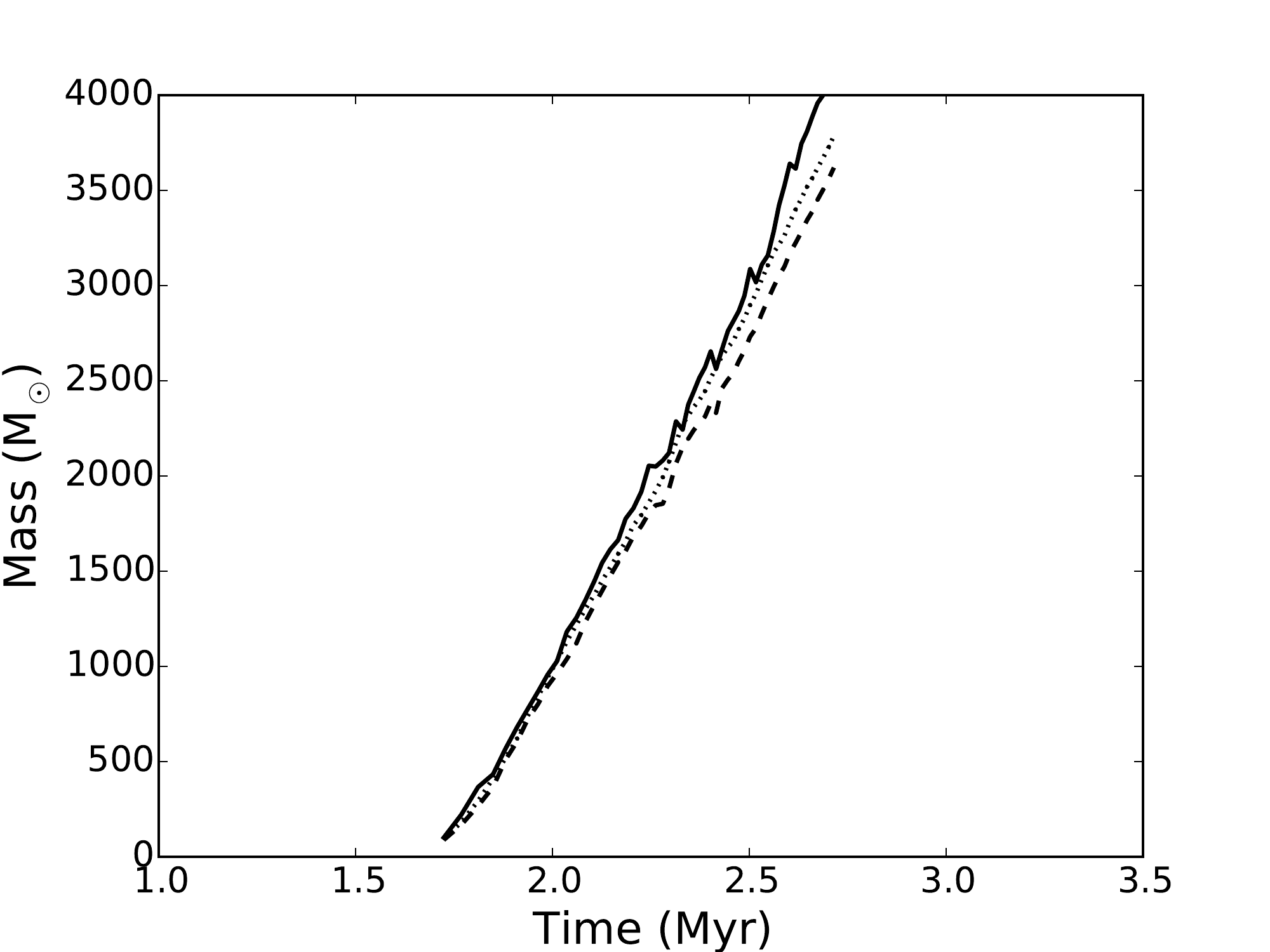}
\end{subfigure}
\begin{subfigure}{.35\textwidth}
\includegraphics[width=\textwidth]{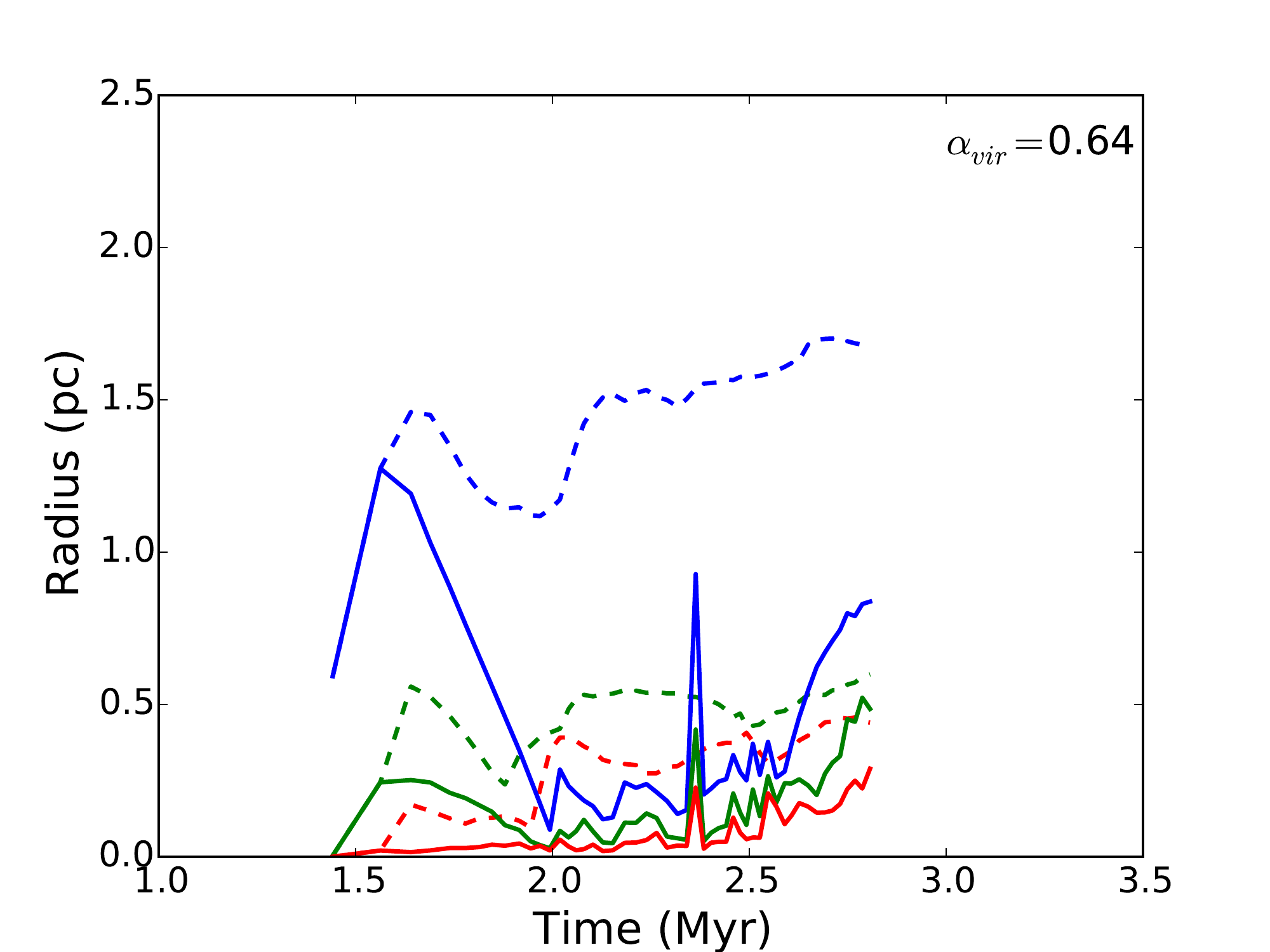}
\end{subfigure}
\begin{subfigure}{.35\textwidth}
\includegraphics[width=\textwidth]{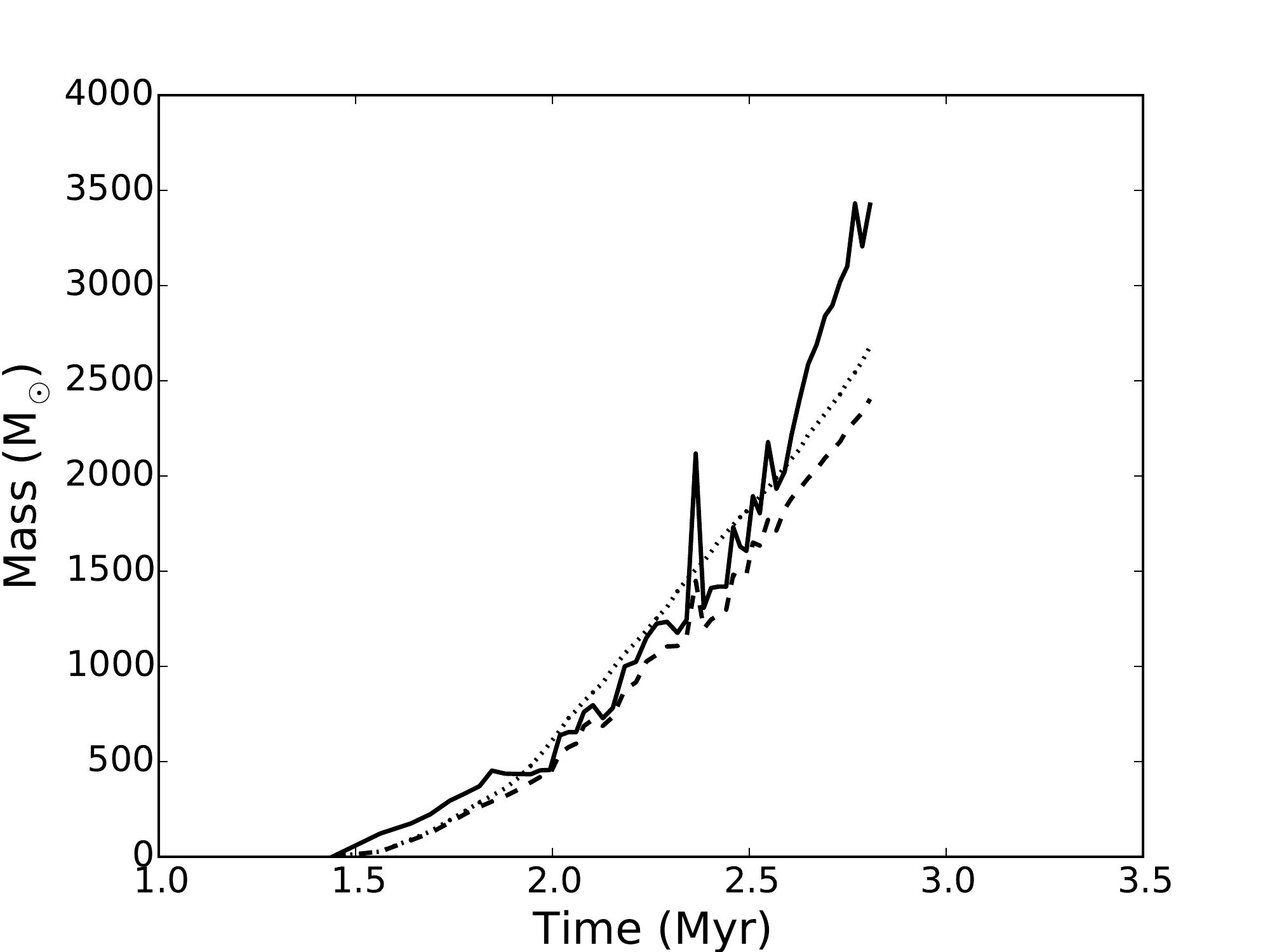}
\end{subfigure}
\begin{subfigure}{.35\textwidth}
\includegraphics[width=\textwidth]{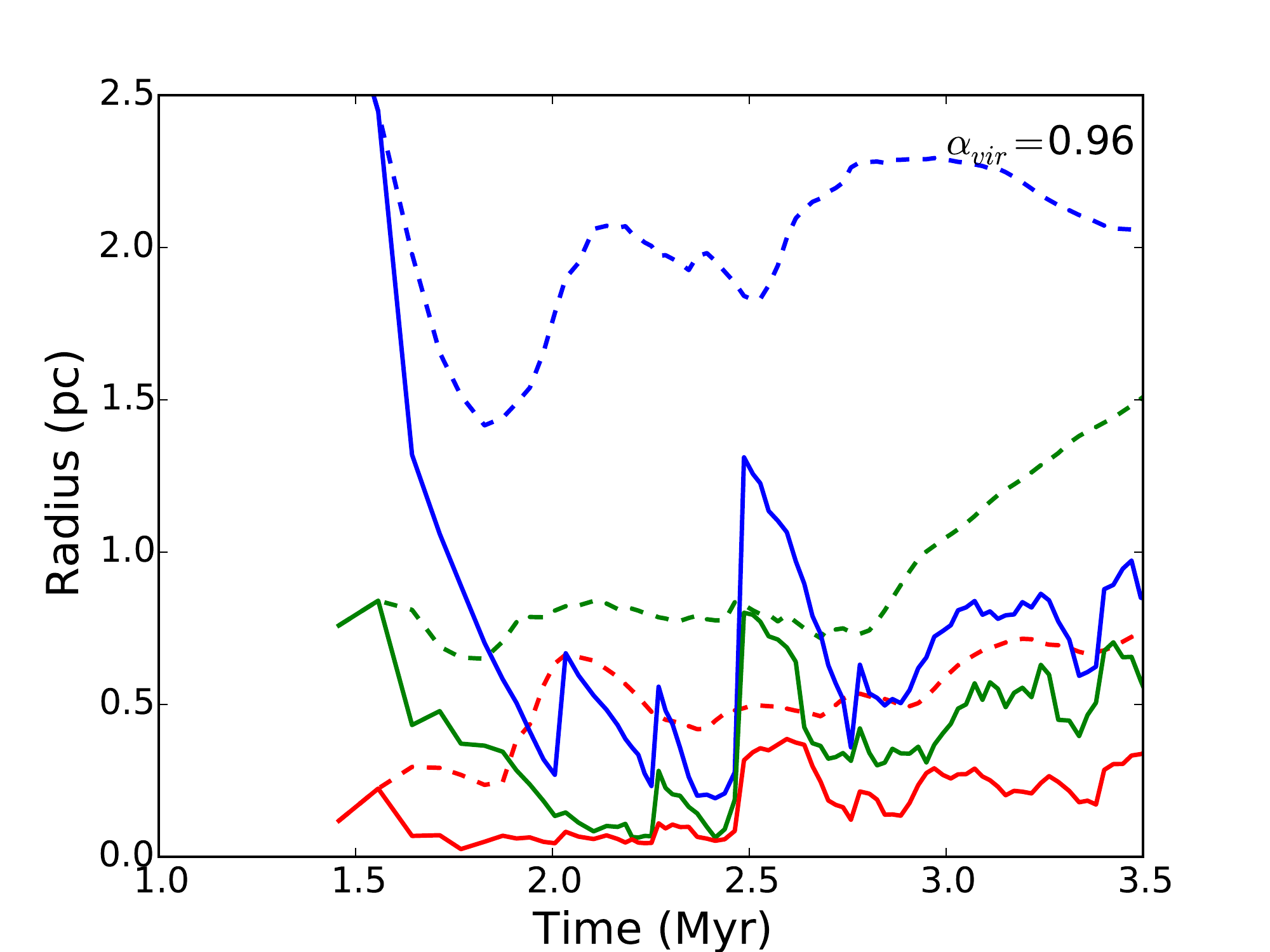}
\end{subfigure}
\begin{subfigure}{.35\textwidth}
\includegraphics[width=\textwidth]{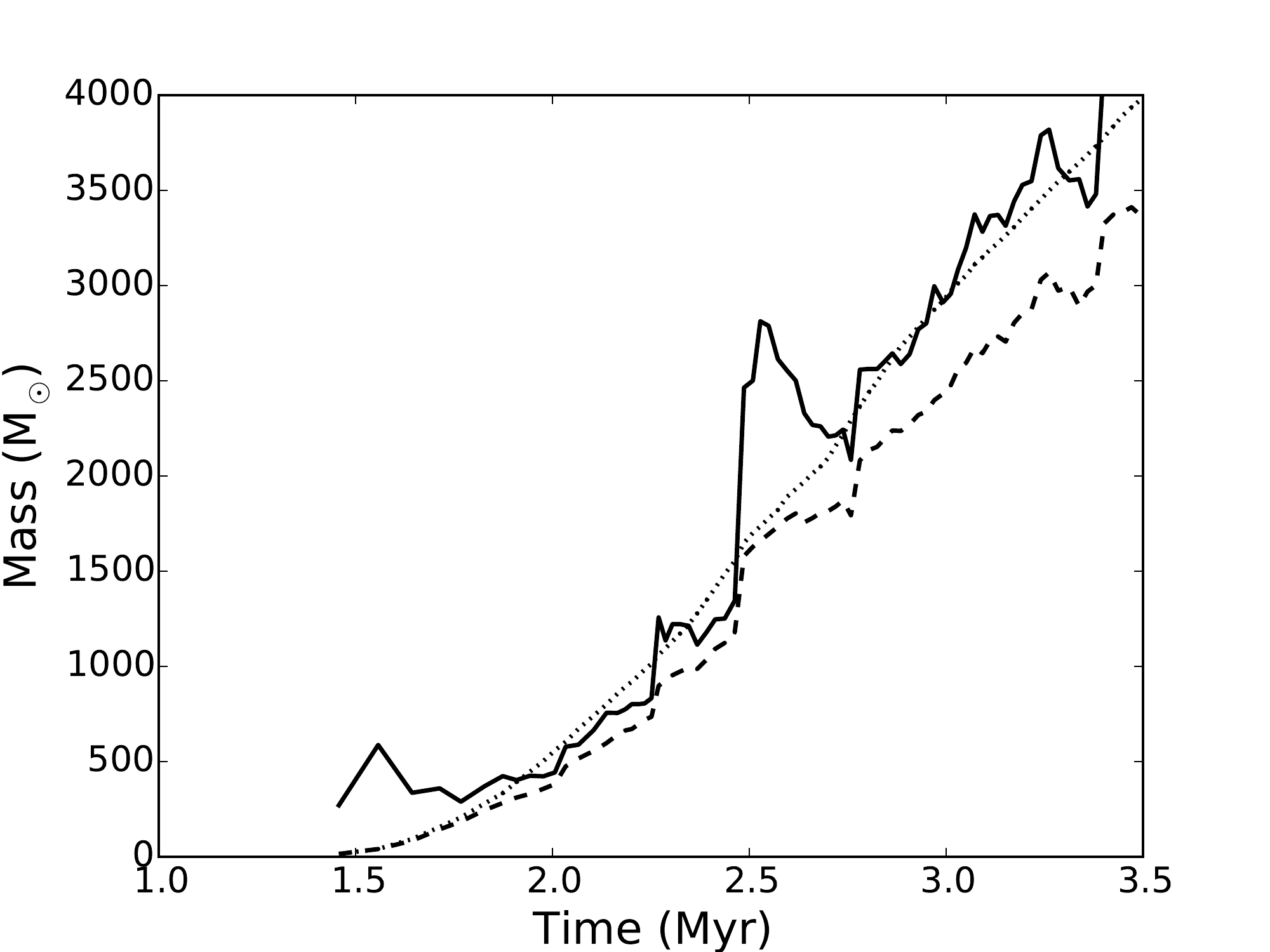}
\end{subfigure}
\begin{subfigure}{.35\textwidth}
\includegraphics[width=\textwidth]{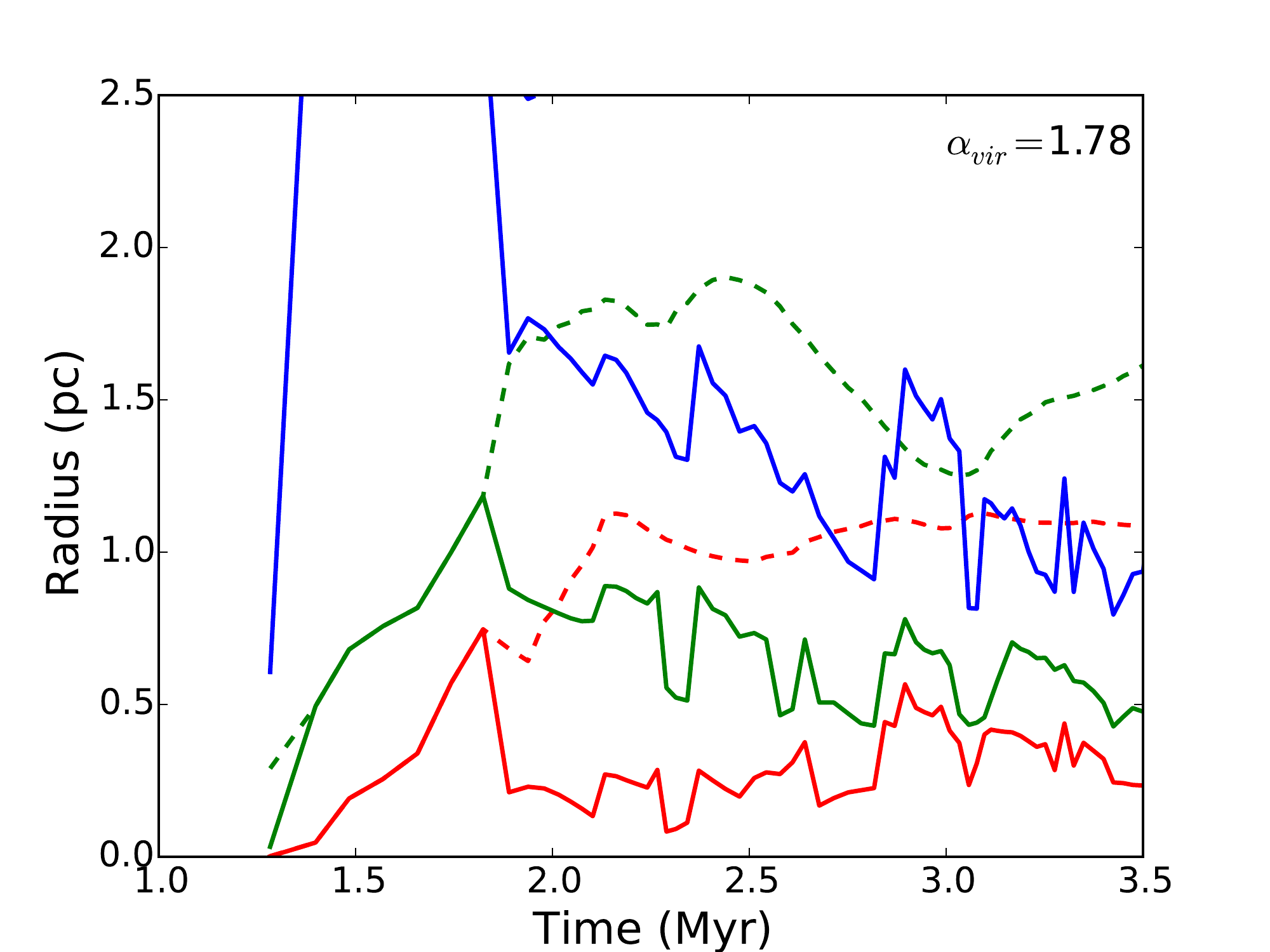}
\end{subfigure}
\begin{subfigure}{.35\textwidth}
\includegraphics[width=\textwidth]{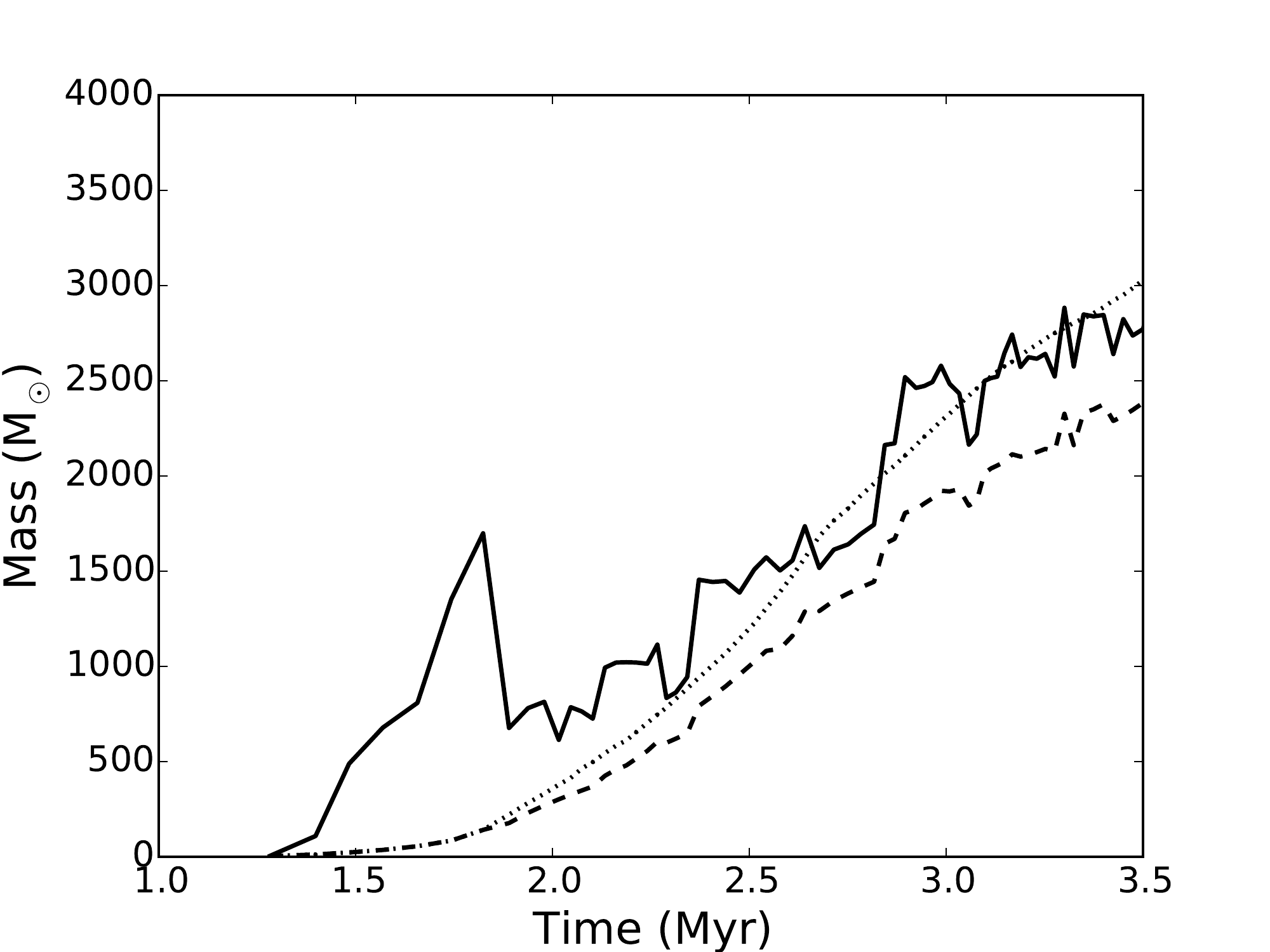}
\end{subfigure}
\caption{The time evolution of the three semi-axes of the cluster determined with sink particles on the left column. 
The dashed curves and solid curves represent the values calculated with all sinks, 
and the reduced values with distant sinks omitted, respectively. 
The sink mass (dashed) and total mass (solid) inside cluster region defined with sink distributions plotted against time on the right column. 
The dotted line represents the sink mass before sink removal. 
Removing the sink particles far away from the center drastically reduces the size while only mildly decreasing the total mass.
From top to bottom are runs A, B, C, and D with increasing levels of turbulence.
}
\label{rt_mt_sinks}
\end{figure*}

From a first calculation (see dashed lines in Fig. \ref{rt_mt_sinks}), 
it is clear that the sink particles are very extendedly distributed,
and that the concentration in certain directions gives a relative large eigenvalue compared to the other two.
However, as we saw before, some sink particles are forming in filaments feeding the cluster 
(see Fig. \ref{simu_3_col}), 
which is consistent with the previous calculations, and therefore are not located inside the cluster. 
Furthermore there are also sinks which are ejected from the cluster
through N-body interactions. 
These sinks, however, should not be taken into account when studying the cluster itself.
We therefore refine the calculation by omitting the sink particles having distances larger than the largest semi-axis from the first calculation and repeating the same procedure.
This operation is not always converging, 
we therefore search for convergence by increasing $\beta$ starting from unity.
When $\beta$ is large enough, 
convergence is reached in just a couple of iterations.
We employ the smallest $\beta$ value which gives convergence, 
which goes up to about 1.5 at largest (implying that $\beta$ is always between 1 and 1.5).
This yields a cluster which is more spherical and stays roughly constant in size as the total sink mass increases.

The evolution of the sink cluster radius and mass are plotted in Fig. \ref{rt_mt_sinks}. 
The total mass of all sinks is plotted in dotted curves while that of those inside the cluster is plotted with dashed curves. 
The total mass including the gas is also plotted in solid curves.
At early times right after the formation of the first sinks, 
the inferred cluster size fluctuates and could go to very large values. 
These are possibly the cases where very few sinks, which form in the surrounding filaments, exist, 
and therefore the algorithm fails. 
The selection of clustered sinks is not always robust, 
and the inferred size shows spiky fluctuations despite the generally stable trend it exhibits. 
However, the significantly reduced size and the relatively mild decrease in mass suggest that the distant sinks which make up only a small mass fraction are effectively rejected. 

\subsubsection{Minimal spanning tree}
It is instructive to confront these results with the ones obtained using another method.
The minimal spanning tree method (MST) \citep{Gower69,Cartwright04} is also applied to examine the selection of sinks belonging to the cluster.
The MST is obtained by linking all the points in a cluster with a tree-like structure,
while minimizing the sum of the length of the links without forming loops.
One way to find the MST is to start with two closest points,
and one shortest branch is added each time to link one more point to whichever point in the group.
A critical value $L_\mathrm{crit}$, the longest branch length allowed inside a cluster,
is used to define the members of a cluster.
The method used by \citet{Gutermuth09} to determine $L_\mathrm{crit}$ with the branch length cumulative distribution function does not work well with our dataset, 
since there are many closely located sinks, making the steep-slope segment extremely steep and thus rejecting more than half of the sinks. 
If we plot the radius calculated for the selected set of sinks as function of $L_\mathrm{crit}$,
it could be seen that with $L_\mathrm{crit}$ increasing, there are abrupt jumps of the cluster size.
These events indicate that a previously considered distant sink or group of sinks is included into the cluster,
and this distance is comparable with respect to the original cluster size. 
This picture is constant with the cluster structure being fractal. 

\begin{figure}[]
\includegraphics[width=0.5\textwidth]{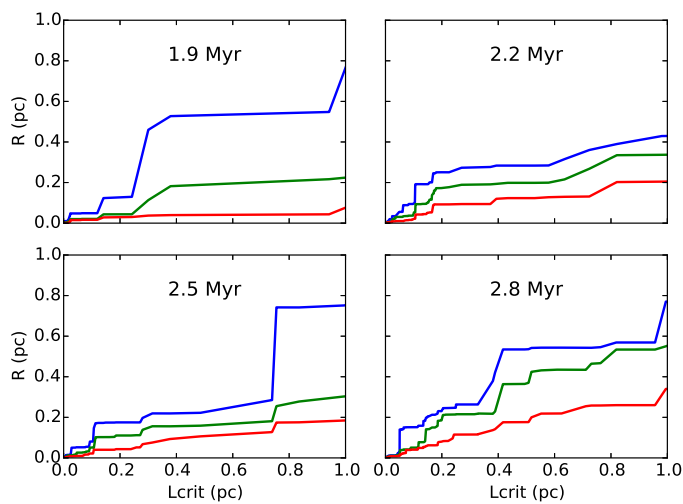}
\caption{The three semi-axis calculated from the rotational inertia matrix with the assumption of uniform mass distribution, plotted against $L_\mathrm{crit}$, the maximal allowed distance linking a member to the cluster. Examples are shown for four time steps of run B.}
\label{MST}
\end{figure}

Some examples of the MST radius-$L_\mathrm{crit}$ plot are shown in Fig. \ref{MST}, several plateaux appear, 
which all indicate possible cluster radius.
This is compatible with the fluctuations of radius defined with the previous method. 
Attempts have been made to select the plateau of which the rotational axis aligns the best with the minor axis,
while the optimal match usually occurs for a very small group of sinks.
Therefore, an automatic determination of cluster member selection is not straightforward.
Nonetheless, most of the time we see one of the plateaux corresponding very well to the radius calculated with the relatively simple sink selection we proposed.
Consequently, we conclude that using a simple selection criteria such as the one we presented to characterize the cluster leads to reasonable results.

\subsubsection{Other methods}

We also tried inferring the sink cluster radius according to the rms distance to the center weighted by local density proposed by \citet{Casertano85}. 
The local density (number or mass) is calculated for each sink considering the $i$th nearest neighbor, and the cluster center is also determined by the density-weighted position. 
The inferred radius is typically less than $10^{-2}$ pc and is very dependent of the number of neighbors used. 
This comes probably from the same problem mentioned in the previous paragraph that we have sink particles very close to each other at the central part of the cluster and therefore creating extremely high local densities. 
The local density could go up to several orders of magnitude contrast for sinks at different locations, 
and the central sinks are thus heavily weighted, giving small radius. 
Slightly differently, \citet{Portegies10} suggested using the squared density as the weight. 
In our case, this only make the problem more serious. 

Another method using the local density is also looked into. 
A neighborhood radius is first determined from the distribution of the sink distances. 
A sink is included into the cluster if the number of sinks in its neighborhood exceed the threshold, 
which comes from the level of significance required for detection. 
The cluster size is than calculated with the minimal convex hull which contains all the cluster members (Joncour in prep.). 
This method detects the departure from a random distribution and finds the over-dense region. 
Again, due to the characteristics of our sink distribution, 
the treatment of multiple systems may also play an important role (internal communication), 
and special care needs to be taken when selecting the parameters.  
With properly selected parameters, the sink cluster radius is similar to what we found with the inertia momentum.

The above described methods set out from observational points of view, 
with the goal of defining a local concentration and its extension. 
As we discussed, the definition of a cluster is not trivial and depends a lot on the parameters used. 
Therefore there always lie some uncertainties in its determination. 
Our simulations do not have sufficient resolution to resolve individual stars, 
therefore methods using local properties tend to work less well. 
The cluster inferred by using the moment of inertia captures the mass distribution at the cluster scale, 
and is more pertinent at least for consistent comparison among simulations.

\subsection{The cluster mass-size relation}

\begin{figure}[h!]
\centering
\begin{subfigure}{0.5\textwidth}
\includegraphics[trim=80 10 100 10,clip,width=\textwidth]{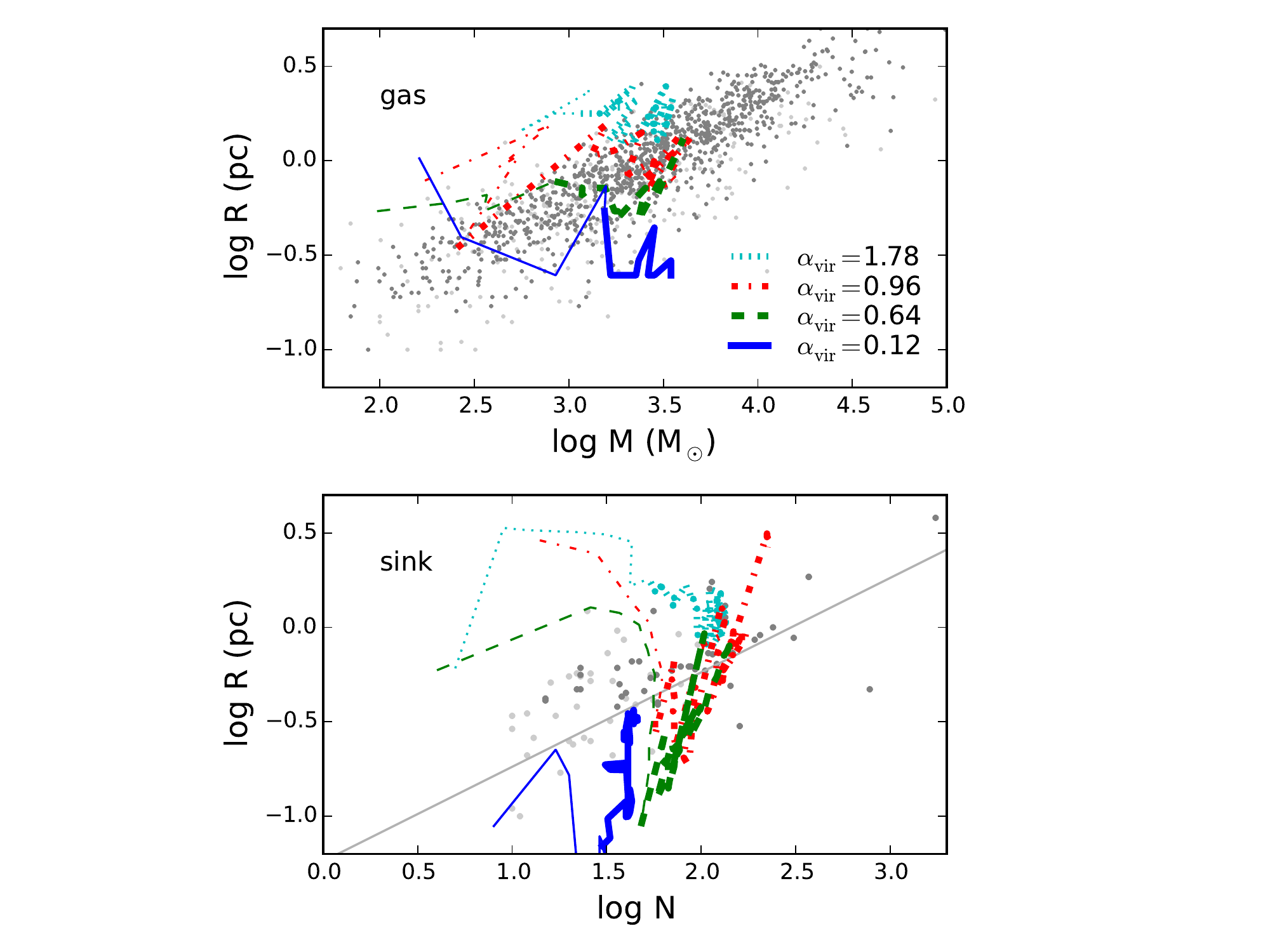}
\end{subfigure}
\caption{Upper panel: Gaseous proto-cluster mass-size relation defined with gas kinematics over-plotted with measurements of star-forming clumps summarized by \citet{Fall10} (light grey dots) and \citet{Urquhart14} (dark grey dots). 
The total mass of the cluster is used (gas plus sinks). 
Lower panel: sink  cluster number-size relation over-plotted with embedded cluster concluded by \citet{Adams06} (dark grey dots) and \citet{Gutermuth09} (light grey dots).
In grey line is the number-size relation $R ~(\mathrm{pc}) \propto \left(N/300\right)^{1/2}$ by \citet{Adams06}.
Simulations with four different levels of turbulent support are plotted. 
The blue solid, green dashed, red dot-dash, cyan dotted curves have virial parameters $\alpha_\mathrm{vir}=$ 0.12, 0.64, 0.96, 1.78 respectively. 
The curves represent the time sequence. As the proto-clusters accrete, they gain in mass and move from left to right in the figure. 
Time steps before 2 Myr are plotted with thinner lines.}
\label{virialsimu}
\end{figure}

The mass-size relation is one of the important characteristics of the proto-clusters that could be compared to observations.
In Fig. \ref{virialsimu}, 
the mass-size relations of gaseous proto-cluster defined with gas kinematics are over-plotted with observations of star-forming clumps \citep{Fall10, Urquhart14}, 
and the number-size relations of cluster obtained from sink distributions are over-plotted with embedded cluster observations \citep{Adams06, Gutermuth09}. 
Early time steps (before 2 Myr) are plotted with thinner lines.
As the gaseous proto-cluster accretes mass,
it arrives on the observed sequence.
The larger the turbulence in the parent cloud, 
the more expanded the proto-cluster since the mass feeding the proto-cluster has higher kinetic energy and the rotation is also more important.
For the sake of comparison with observations, we also show the relation obtained by \citet{Adams06} $R ~(\mathrm{pc}) \propto \left(N/300\right)^{1/2}$. 
The sink clusters come close to this sequence after sufficient evolution. 
At later stage, they form less new sinks and start increasing in radius. 
Further discussion is however needed here as our sinks are  massive and do not represent individual stars. 
The number-size (or mass-size) relation inferred with sink particles is thus biased and not exactly representative of the embedded cluster itself. 
The absence of feedback in our simulations translates into over-formation of sink particles and their over-accretion (over-estimation of star number and mass), 
and these sinks represent prestellar cores which might form stars inside (under-estimation of star number) with some efficiency (over-estimation of star mass). 
The overall effect is possibly a shift of the number-size relation in the lower panel of Fig. \ref{virialsimu} towards the right, 
with the upper limit 20 being the ratio between averaged sink mass (10 M$_\odot$) and averaged stellar mass (0.5 M$_\odot$).
On the other hand, 
the cluster is a hierarchical structure. 
The size is very dependent of the membership selection and there could easily be a factor 2 difference in radius (see Figs. \ref{rt_mt_sinks}, \ref{MST}) as consequence of different sensitivity and criteria used, 
which would lead to a factor 10 in density uncertainties. 
The expansion of cluster after gas removal by stellar feedback is not included in our simulations, 
giving another possible under-estimation of the cluster radius. 
Considering all these uncertainties, 
we cannot readily reproduce the number-size relationship for embedded clusters, 
but they should at least not be too far away from the observed sequence. 
Observationally speaking, 
\citet{Pfalzner16} plotted the number-size relation for different star-forming regions and found that, 
though following the same trend, 
this relation could have more than 20 times scatter in number inside different regions (their Fig. 3). 
The cluster definition and the distance of the object could play important roles. 
On the other hand we see that the sink cluster radius does not change very much with time, 
at least telling us that at the early stage of stellar cluster formation, 
they should still somewhat be correlated to and regulated by its natal gas. 
We stress that according to our analysis, 
the link between the gaseous proto-cluster radius and that of the embedded cluster is not a trivial matter. 
Therefore the 20 \% efficiency that is usually inferred should be regarded with great care.

\begin{figure}[h!]
\centering
\begin{subfigure}{0.5\textwidth}
\includegraphics[trim=20 10 20 15,clip,width=\textwidth]{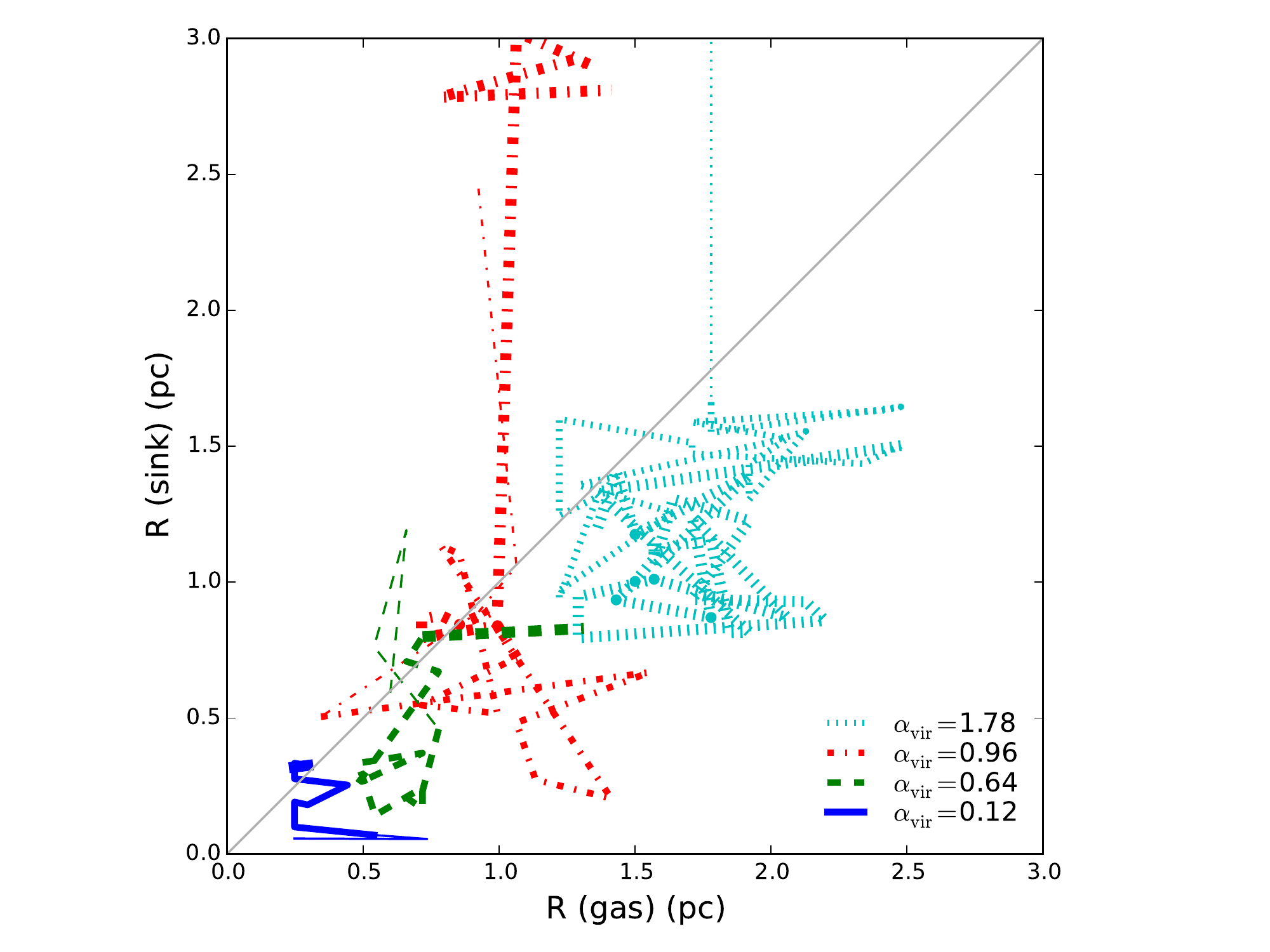}
\end{subfigure}
\caption{The radius of sink clusters plotted against radius of gas proto-clusters for 4 runs. 
The relations are plotted with thinner lines for time before 2 Myr, and with thicker lines after 3 Myr.
The color coding is the same as that in Fig. \ref{virialsimu}.
The gas and sink cluster sizes show good correlation in general, 
while the sink cluster size is slightly smaller than that of the gas cluster.}
\label{Rg_Rs}
\end{figure}

\begin{figure}[h!]
\vspace{-1.\baselineskip}
\centering
\includegraphics[width=0.5\textwidth]{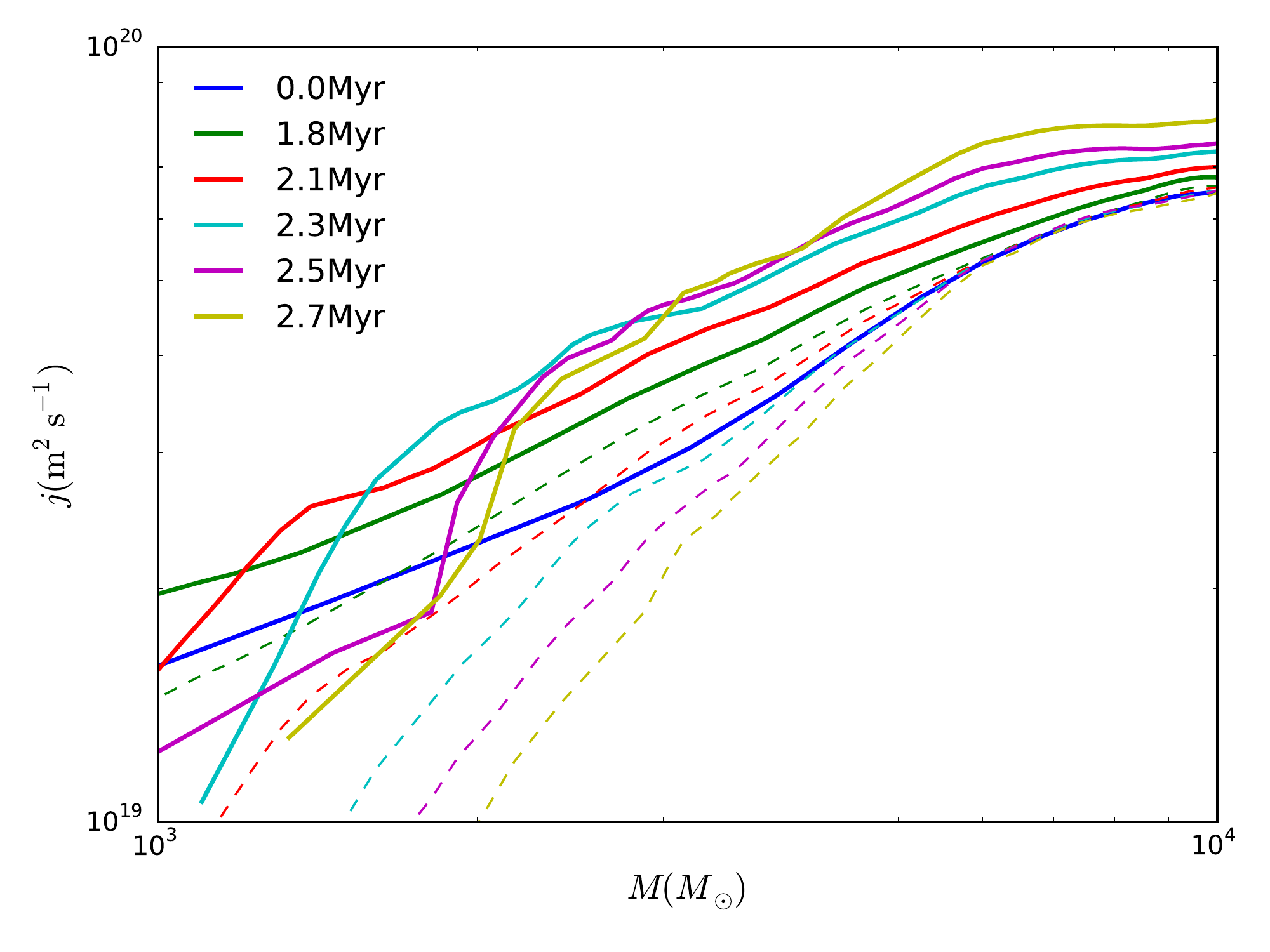} 
\vspace{-2.5\baselineskip}
\caption{The specific angular momentum plotted against mass contained inside ellipsoids of varying radius at a series of time steps of run B.
Solid curves represent the specific angular momentum of gas.
Dashed curves represent the averaged value considering gas and sinks at the same time.
The specific angular momentum of gas is more or less conserved during the collapse, 
which justifies our simplification,
while the sink particles exhibit a loss in angular momentum. 
}
\label{J_M_simu}
\end{figure}

We compare the cluster size inferred with gas and sinks in Fig. \ref{Rg_Rs}.
The sink clusters generally have smaller size than  the gas clusters. 
This implies that the stars are possibly formed in the inner region of the gaseous proto-cluster. 
The relations are plotted with line thickness that increases with time. 
We can see with both the green and red curves that at early time the sink cluster radius is large probably due to bad definition of cluster with very few sinks. 
Once the cluster becomes steadily established (after 2 Myr), they show a radius smaller than the gaseous proto-cluster size.
At even later time (after 3 Myr with the thickest lines), the sink cluster radius shows a growth with respect to the gas radius.
This might be a sign of the dynamical decoupling and the expansion of the stellar cluster. 
One evidence of the sink-gas decoupling is indeed shown in Fig. \ref{J_M_simu}, 
where we plot specific angular momentum of gas (solid curves) and gas plus sinks (dashed curves) in concentric ellipsoids against mass at several time steps.
For a fixed mass, the region concerned is shrinking in time under the collapse. 
The solid curves are relatively close to one another, 
which means that in one collapsing mass entity, 
the averaged gas specific angular momentum is not varying significantly in time. 
Moreover, we see that while the total angular momentum (dashed curves) is conserved at large mass ($M \gtrsim 5 \times10^3$ M$_\odot$), 
it decreases at lower mass, indicating that the angular momentum is being transported away. 
We conclude that while the gas conserves more or less its angular momentum during the collapse,
the momentum of the sinks is lost gradually.
\citet{Longmore14} concluded that the gas-star co-evolution cannot continue over a dynamical time,
on one hand because the gas is dissipative while stars are ballistic,
on the other hand because stars form from gravitational collapse of gas and create for themselves a local gas-free environment.
Also mentioned by \citet{Bate03}, the gas-star interaction plays a very important role in regulating the star formation.
\citet{Parker15} perform \cal{Q}-parameter analysis \citep{Cartwright04} of gas and stars in simulations and suggest that their spatial distributions are not trivially linked.
This sheds light on the importance of correctly following both gas and particle dynamics in simulations and the necessity of pertinent theories  of early stellar cluster formation.

To conclude,  our simulations successfully reproduce the mass-size regions that we define as gaseous proto-clusters
provided the virial parameter $\alpha_\mathrm{vir}$ is large enough.
When the cloud is too weakly supported (small $\alpha_\mathrm{vir}$),
the gaseous proto-cluster is strongly accreting and has a small radius
and this should probably not be compared to the observations.
For other runs of clouds with $\alpha_\mathrm{vir}$ close to unity,
a good agreement of the mass-size relation between observations and simulations is reached.
As gaseous proto-clusters accrete mass, they arrive on the observed sequence, 
which implies very likely a stationary and quasi-equilibrium state. 

Although we have performed simulations of $10^4$ M$_\odot$ clouds, 
giving gaseous proto-clusters of a few thousand solar masses, 
the gas is almost isothermal at the cloud and proto-cluster scale, 
and the results could then be rescaled to the observed mass range. 
The simulations are parametrized by  two non-dimensional numbers: 
\begin{align}
\alpha_\mathrm{vir} \propto {\sigma^2 R \over M}  ~~~ \mathrm{and}~~~~
\mathcal{M} = {\sigma \over c_\mathrm{s}}
\end{align}
If we rescale the molecular cloud in the simulation to different mass, size, and temperature such that
\begin{align}
M^\prime = A M ~~~,~~~
R^\prime = A^{1\over 3-\gamma} R  ~~~ \mathrm{and}~~~~
c_\mathrm{s}^\prime = A^{2-\gamma\over 6-2\gamma},
\end{align}
where $\gamma = 1 \text{ or } 0.7$ is the exponent in the Larson's relation $\rho \propto R^{-\gamma}$ \citep{Larson81,Falgarone04, Falgarone09,HF12, Lombardi10} and $A$ is a scaling factor. 
With this rescaling, $\alpha_\mathrm{vir}$ and $\mathcal{M}$ stay unchanged, 
and the new cloud also follows the Larson's relation. 
The gaseous proto-cluster inside the cloud is therefore rescaled following 
$R_\ast \propto M_\ast^{1\over 3-\gamma} \sim  M_\ast^{0.5} $. 
This implies that gaseous proto-clusters form inside molecular clouds which follow Larson relations
follow the relation reported by \citet{Fall10} and \citet{Urquhart14}. 
Indeed in Paper II, 
an analytical model is developed to account for the mass-size relation of gaseous proto-clusters of any mass.
Given the turbulence level and the mass of the parent cloud,
the model predicts the mass, size, aspect ratio, and velocity dispersion of the gaseous proto-cluster. 
As long as the conditions that stellar feedback is not very important stays valid, 
our simulations and analytical model match with the observations of low mass gaseous proto-clusters.

Stars form out of the gas reservoir in the gaseous proto-cluster. 
Properties possessed by the gas should be to some extent passed down to the stars.
Once formed, the stars, however, start to dynamically depart from the natal gas.
Links between the proto-cluster gas and the stellar cluster should be carefully made to yield comprehensive comparisons and shed light on their different dynamical properties.
This is seen in the mass-size relations derived from observations of stars in embedded cluster \citep{Adams06} and gas of star-forming clumps \citep{Fall10,Urquhart14}.
Though showing the same trend,
they exhibit a shift in absolute value \citep{Pfalzner16}.
If we assume $0.5$ M$_\odot$ averaged mass for the stars inside the clusters, 
a star formation efficiency less than $10\%$ is required to reconcile the two relations given that the cluster size does not evolve. 
Alternatively, this could also be explained by an expansion after the cluster formation. 
We inferred the cluster size for gas and sinks, which allows a comparison between the two and with observations as well.


\section{Internal properties of the proto-cluster}
The clusters in the simulations having been defined,
we can calculate its mass, size, angular momentum, velocity dispersion and various quantities.
In this section, we perform some analysis of the proto-cluster properties
since they are important for understanding in which conditions stars form.

\subsection{Energies equilibrium}
\begin{figure}[]
\includegraphics[width=0.5\textwidth]{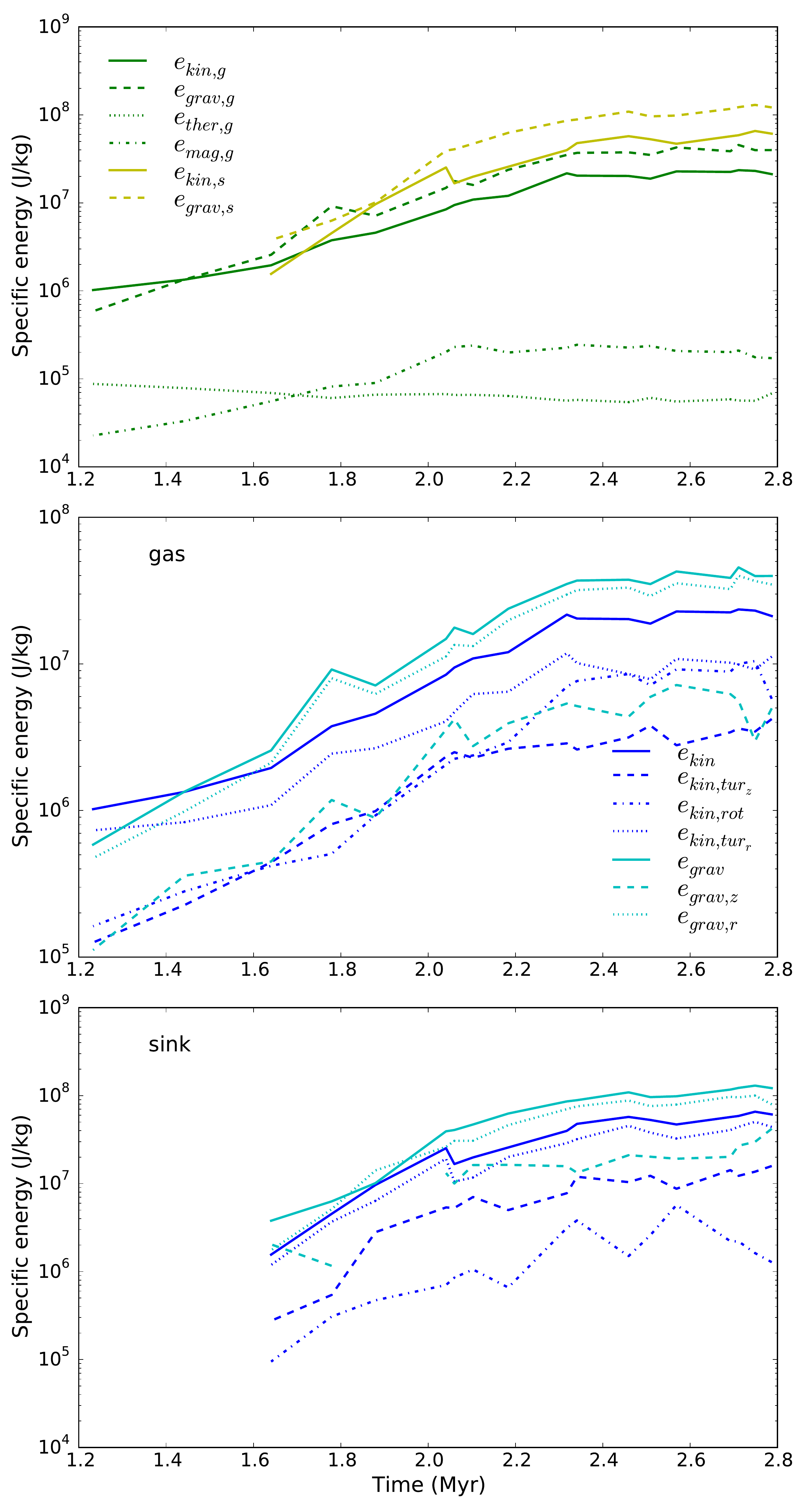}
\caption{Top: different forms of specific energy of the cluster defined with gas kinematics plotted against time.
The gas component is plotted in green, and the sinks in yellow.
The solid, dashed, dotted, and dash-dot lines represent kinetic, gravitational, thermal, and magnetic energies respectively.
The gravitational energy is plotted in absolute value.
Middle and bottom: kinetic (blue) and gravitational energy (cyan) decomposition of the gas component (middle panel) and the sink component (bottom panel) of the proto-cluster, plotted in specific values. 
Total energy, energy along $z$-axis, rotational energy, and energy perpendicular to $z$-axis
are plotted with solid lines, dashed lines, dash-dot lines, and dotted lines respectively.
}
\label{E_time}
\end{figure}
Once the proto-cluster is identified,
we proceed to calculate the energies in different forms inside this region and examine the balance.
Since we are discussing the gas-dominated early phase of cluster formation,
we use the proto-clusters previously defined with gas kinematics,
which indeed show good correspondence with observations.
The specific kinetic, gravitational, thermal,
and magnetic energies are plotted at several time steps in the top panel of Fig. \ref{E_time}.
The gravitational energy is negative and is plotted in absolute value.
Here we show only the run with $t_\mathrm{ff}/t_\mathrm{vct} = 0.9$ since all the runs exhibit similar trends.
Results of other runs are shown in appendix \ref{E_ana_others}.

For the gas component (green), 
the thermal and magnetic energies are a few percents of the gravitational energy, 
consistent with that initialized in the parent cloud, 
and therefore do not contribute much to supporting against self-gravity at the cluster scale.
The specific thermal energy stays roughly constant and decreases mildly,
indicating a slight temperature decrease due to density increase.
The specific magnetic energy increases slightly as a result of magnetic flux concentration.
On the other hand, the kinetic energy, which has contributions from turbulence and rotation,
acts largely to support against gravitational contraction.
The cluster satisfies $2E_\mathrm{kin,g} + E_\mathrm{grav,g} \approx 0$, 
indicating the gas component is in virial equilibrium while the cluster is accreting mass.
The particle component (yellow), on the other hand,
begins to have higher specific kinetic and gravitational energies a short while after the sinks start forming.
This implies that the sink particles are more centrally concentrated than the gas and lie in a deeper potential well, 
which is coherent with our previous finding that the radius determined with sinks is smaller (Fig. \ref{Rg_Rs}). 
This is compatible of the discovery by \citet{Bate03} that the stars have velocity dispersion 3 times larger than that of the gas.
It is remarkable that the sinks are bound almost at virial and satisfy the relation $2E_\mathrm{kin,s} + E_\mathrm{grav,s} \approx 0$,
indicating that the cluster energy properties are largely inherited from its natal gas and is determined at the early formation stage.
\citet{Walker16} suggested that clusters form very likely in a conveyor-belt mode, in which the core of the star-forming cloud accumulates mass at the same time as stars form. This is perfectly in coherence with our simulation results that stars form in the dense gaseous proto-cluster, of which the mass and energy is sustained by accretion.

\subsection{Rotation and turbulence anisotropy}
The rotation makes up an important part of the kinetic energy of the proto-cluster, 
and a flattened form is thus a natural consequence. 
We estimate the rotational energy of the proto-cluster and separate it from its turbulent counterpart.
This allows us to compare the proportions of rotational and turbulent energies 
and also examine whether turbulence is isotropic.
The rotational energy is estimated as $E_\mathrm{kin,rot}={1 \over 2} J I^{-1} J$,
where $J$ is the total angular momentum vector of the cluster, 
and $I$ is its rotational inertia matrix.
This  allows us to  eliminate turbulence by summing up various motions.
There  remain uncertainties, in particular since it does not take into account the differential rotation
 this is probably an underestimation.
The one dimensional turbulent energy along the rotational axis of the cluster $E_{\mathrm{kin},z}$ is also calculated.
The turbulent energy perpendicular to the rotational axis could be thus estimated as $E_{\mathrm{kin},r} = E_\mathrm{kin} - E_\mathrm{kin,rot} - E_{\mathrm{kin},z}$.
We display the energies for the run with $t_\mathrm{ff}/t_\mathrm{vct} = 0.9$. 
Other results are shown in appendix \ref{E_ana_others}.
In the middle panel of Fig. \ref{E_time}, we plot with blue color the total kinetic energy in solid line, rotational energy in dot-dash line, 
the turbulence parallel and perpendicular to the rotational axis in dashed and dotted line respectively for the gas component.

The proportion of rotational energy becomes more and more important as the gaseous proto-cluster accretes,
while that of $E_{\mathrm{kin},z}$ is slightly decreasing.
The rotational energy grows to become comparable to the turbulent energy in the same plane and makes up nearly a third of the total kinetic energy.
The turbulence shows anisotropy since $E_{\mathrm{kin},r}/E_{\mathrm{kin},z}>2$,
although  there remain uncertainties in the estimation of the rotation.
The rotation flattens the proto-cluster,
and this anisotropy in geometry thus in turn makes the kinetic energy distribution anisotropic.

We also plot in cyan the gravitational energy of the cluster ($E_\mathrm{grav}$ solid line).
The gravitational energy is calculated by integrating over the cluster volume the dot product of the gravity and the position vector with respect to cluster center (as is done in the virial theorem).
With this definition, 
we can separate the gravitational energy into two contributions by calculating respectively with the gravitational acceleration parallel ($E_{\mathrm{grav},z}$ dashed line) and perpendicular ($E_{\mathrm{grav},r}$ dotted line) to the cluster minor axis 
(see paper II for definitions). 
This points out the fact that not only the proto-cluster is generally in virial equilibrium,
it also satisfies a modified virial theorem which accounts for the two dimensions:
$2E_{\mathrm{kin},z} + E_{\mathrm{grav},z} \approx 0$ and
$2E_{\mathrm{kin},r} + 2E_\mathrm{kin,rot} + E_{\mathrm{grav},r} \approx 0$.
This motivates the decomposition of the virial theorem into two dimensions in the analytical model in Paper II.
The bottom panel of Fig. \ref{E_time} shows the same plot as that in the middle for the sink component of the proto-cluster.
The sink particles show similar trends to the gas in general. 
A very different behavior is that the rotational energy calculated for the whole system is not increasing like that of the gas. 
Moreover, $E_\mathrm{kin,rot} \ll E_{\mathrm{kin}}$, 
indicating that the stars are giving angular momentum to the gas and there is less a general rotation. 

\subsection{Density PDF}
\begin{figure}[]
\includegraphics[trim=0 5 0 35,clip,width=0.5\textwidth]{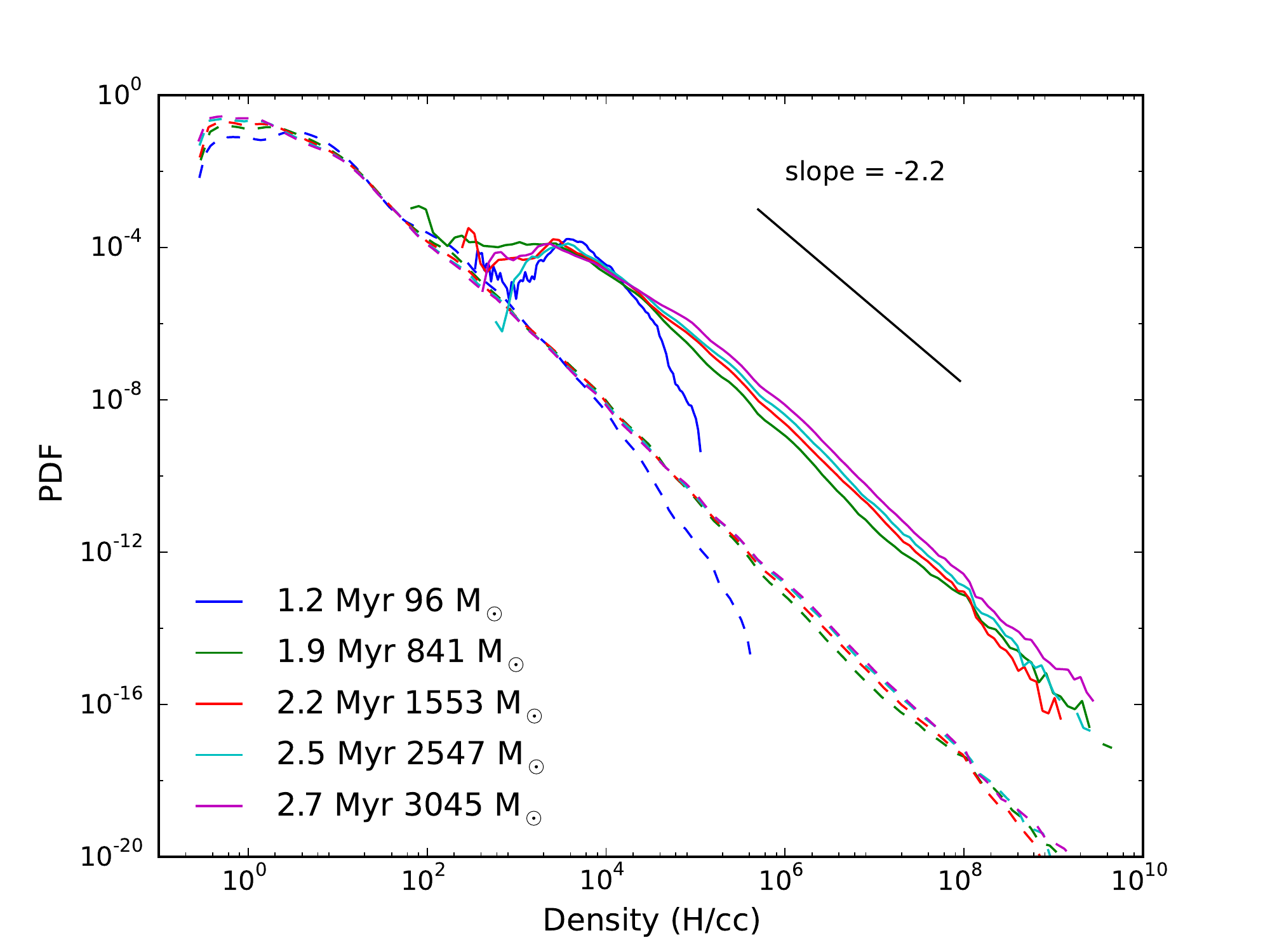}
\caption{PDF at several time steps for run B with $t_\mathrm{ff}/t_\mathrm{vat}=0.9$,
in solid curves for the gaseous proto-cluster,
and in dashed curves for the parent cloud.
The normalization is done such that the integral is equal to one. 
The distribution is close to log-normal at the beginning as a result of turbulence interaction.
The gravity soon dominates and create a prominent power-law tail.
The slope -2.2 is the average of the cluster PDF slopes evaluated at the same density range as the black line except for the first time step.
}
\label{pdf}
\end{figure}
The star formation is a result of collapse induced by local over-density,
therefore essential informations are imprinted in the density probability density function (PDF) of the star-forming environment,
which is in turn a result of interaction between turbulence and gravity.
The density PDF is important for theoretical prediction of the IMF such as that of \citet{Padoan02}, \citet{HC08,Hc09}
and \citet{hopkins12}.
We plot the density PDFs of the gaseous proto-clusters identified in the previous section and that of the parent cloud,
and show that there are indeed fundamental differences.

In Fig. \ref{pdf}, the volume-weighted PDF of the gaseous proto-cluster and that of the parent cloud are plotted at several time steps.
At the beginning of the simulation, 
the PDF is very close to a log-normal distribution, which is a natural consequence of the turbulence,
while the gravity soon dominates at later times.
The gaseous proto-cluster shows the same power-law high mass tail as the parent cloud with a slope $-2.2$,
which is a signature of local gravitational collapse.
This is a common feature of star-forming region as first mentioned observationally by \citet{Kainulainen09}.
The major difference lies in that the gaseous proto-cluster region is denser than the original cloud environment,
with a density peak around $10^4 ~\mathrm{H} ~\mathrm{cc}^{-1}$.
The power-law exponent is slightly steeper than that expected from the self-similar Larson-Penston collapse ($-1.5$) or the expansion wave collapse solution ($-2$)  \citep{Larson69,Penston69,Shu77,Kritsuk11} .
The density PDF of the cloud stays almost unchanged while a gaseous proto-cluster develops within.
This emphasizes the fact that when the stars form,
their environment is already not the same as the cloud in which they reside. 
Inside the gaseous proto-cluster, the gas properties are modified by the interaction of gravity and turbulence.
The gaseous proto-cluster stage should therefore be taken into account when linking the molecular cloud to the stellar cluster.

\subsection{Discussion}
It is still an unanswered question whether the universality of the IMF is a result of similar star formation processes in various kinds of environment,
or a fruit of coincidence by different factors canceling out.
Understanding the underlying mechanisms of star formation is therefore essential.
One key question to answer is whether star formation is dominated by its environment, i.e. the initial gas reservoir available, or by the local gas flow properties.

Stellar clusters are ideal locations to provide statistics on star formation and to study the interaction of stars and their environment. 
Stars form inside the gaseous proto-cluster where the gas is reprocessed during the gravo-turbulent collapse of the parent cloud. 
Compared to a molecular cloud, which follows more or less the Larson relations,
the gaseous proto-cluster environment is indeed different as could be seen with the mass-size relation.
They are 10 times more massive than the clouds at the same size.
The density PDFs also exhibit significant difference, 
which have fundamental influences on the over-density seeding of prestellar cores.
Finding a mapping from the parent cloud to the gaseous proto-cluster would provide a intermediate link between the initial diffuse cloud gas and the dense prestellar cores,
and therefore decoupling the core formation and the initial cloud conditions to some extend if the gaseous proto-cluster properties are more or less general regardless of the parent cloud.

Given our results, we could thus reasonably propose the scenario that 
the gaseous proto-cluster has its turbulence primarily driven by the accretion,
as is a universal process described by \citet{Klessen10,Goldbaum11},
and this turbulence acts as a support against self-gravity,
while at the same time, 
the rotation is important since at this scale the loss of angular momentum by its outward transport remains limited.
It is therefore clear that a primary gas-dominated proto-cluster stage can be defined before cluster formation.
In the companion paper (paper II) we introduce an analytical model,
which describes an accreting ellipsoidal system in virial equilibrium, 
and demonstrate that the gaseous proto-clusters lie on a sequence of global equilibrium as a result of the interplay between turbulence and gravity.
We derive a mapping from the cloud parameters to the gaseous proto-cluster properties.


\section{Conclusions}
Star formation is a hierarchical process which incorporates multiple spatial and temporal scales, 
where the interaction of turbulence and gravity plays an essential role.
In this paper, we focus on the very beginning stage of stellar cluster formation. 
We propose that a primarily structure, the gaseous proto-cluster, is first created out of the collapsing molecular cloud, 
 and it reaches a state of global virial equilibrium as a consequence of the gravo-turbulent interaction.
The gaseous proto-cluster subsequently evolves into the stellar cluster,
which inherits its equilibrium properties from the gas.

A series of collapsing molecular clouds are simulated to form gaseous proto-clusters,
and a sink particle algorithm is used to follow prestellar cores.
The gaseous proto-cluster is clearly identifiable after about 2 Myr, while its boundary is very irregular.
The gas kinematics and sink particle distributions are used to determine the proto-cluster region,
an ellipsoid flattened by its prominent rotation.
We define a measure of infall motion against rotation of the gas as an integrated property,
and infer the gaseous proto-cluster size by distinguishing between the collapsing envelope and the quasi-stationary core.
The sink particle distribution is used to calculate the rotational inertia, 
and in turn yields the cluster size and shape.
The proto-clusters inferred with the aforementioned methods are compared to the observed star-forming clumps and embedded clusters respectively.
Both the gas and sink clusters show stationary behavior in size and are coherent with observations. 
We stress that although the stellar cluster radius and the gaseous proto-cluster radius are correlated, 
their exact values are sensitive to the definition adopted. 
Therefore, this implies that any interpretation in terms of gas removal or efficiency should be taken with care.

From the energy analysis,
the proto-cluster is in virial equilibrium such that the rotation and turbulence support against self-gravity.
As turbulence is driven by the accretion from the collapsing molecular cloud, 
a new balance is established in this emerging entity.
This is obviously prescribed by the properties of the parent cloud and the nature of its collapse. 
The initial turbulence level in the parent cloud is also imprinted in that of the gaseous proto-cluster, 
and determines its size in consequence.
We conclude that the gaseous proto-cluster is indeed an important intermediate step of stellar cluster formation from molecular cloud, 
and that star formation should be studied in this context.
This work does not take into account stellar feedback.
To fully describe the proto-cluster at later times,
a more complete model considering the gas-sink interaction is necessary.


\begin{acknowledgements}
This work was granted access to HPC
   resources of CINES under the allocation x2014047023 made by GENCI (Grand
   Equipement National de Calcul Intensif). 
   This research has received funding from the European Research Council under
   the European Community's Seventh Framework Programme (FP7/2007-2013 Grant
   Agreement no. 306483).
   The authors thank the anonymous referee for the careful reading and useful suggestions. 
\end{acknowledgements}

\appendix
\section{Optimal characterization of the infalling envelope}
\label{Ravg}
\begin{figure*}[]
\centering
\begin{subfigure}{.33\textwidth}
\includegraphics[trim=20 0 40 20,clip,width=\textwidth]{gas_figs/rt_3_res.pdf}\caption{$(R^2H)^{1\over 3}, W_0$}
\end{subfigure}
\begin{subfigure}{.33\textwidth}
\includegraphics[trim=20 0 40 20,clip,width=\textwidth]{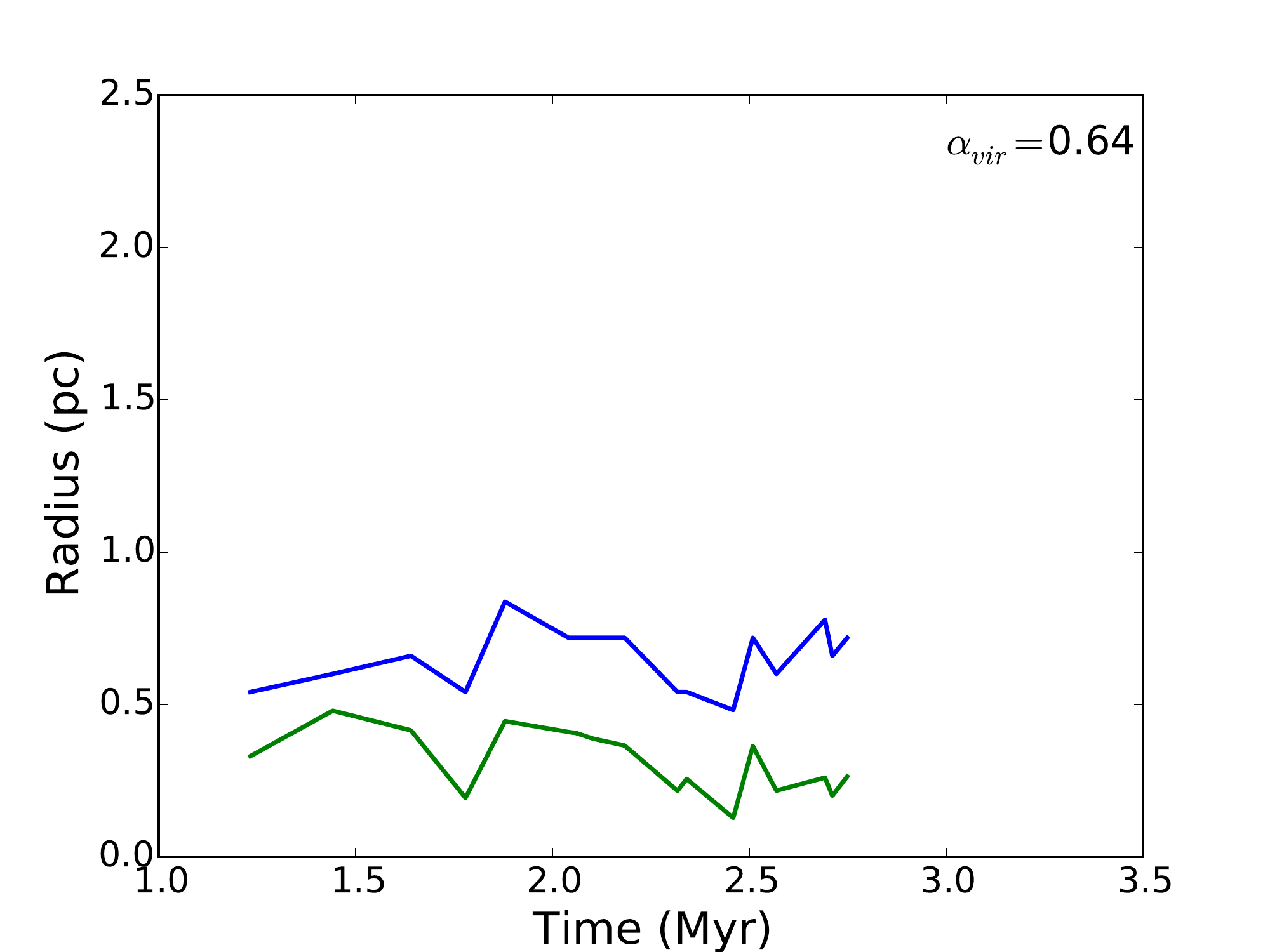}\caption{$(RH)^{1\over 2}, W_0$}
\end{subfigure}
\begin{subfigure}{.33\textwidth}
\includegraphics[trim=20 0 40 20,clip,width=\textwidth]{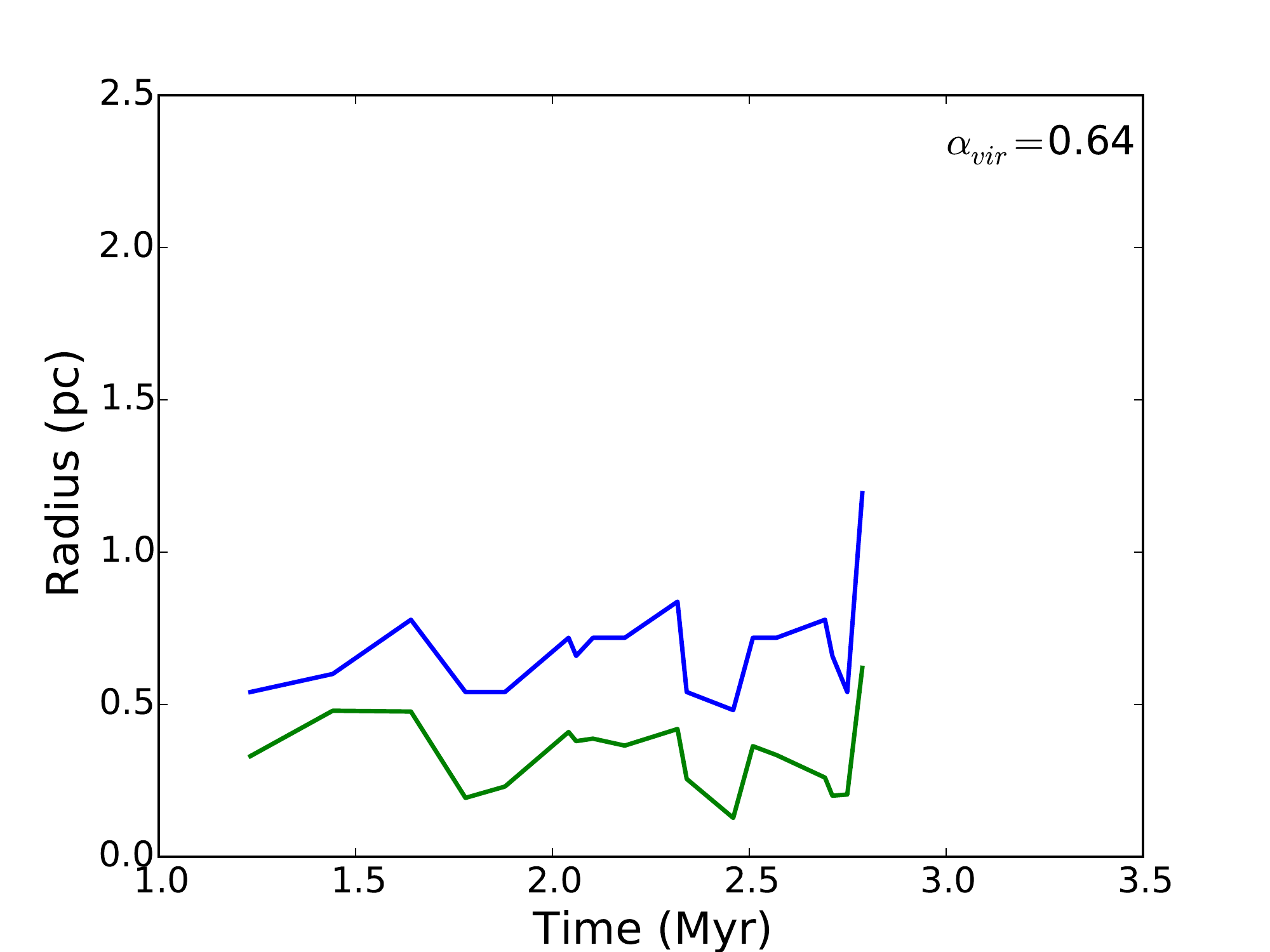}\caption{$R, W_0$}
\end{subfigure}
\begin{subfigure}{.33\textwidth}
\includegraphics[trim=20 0 40 20,clip,width=\textwidth]{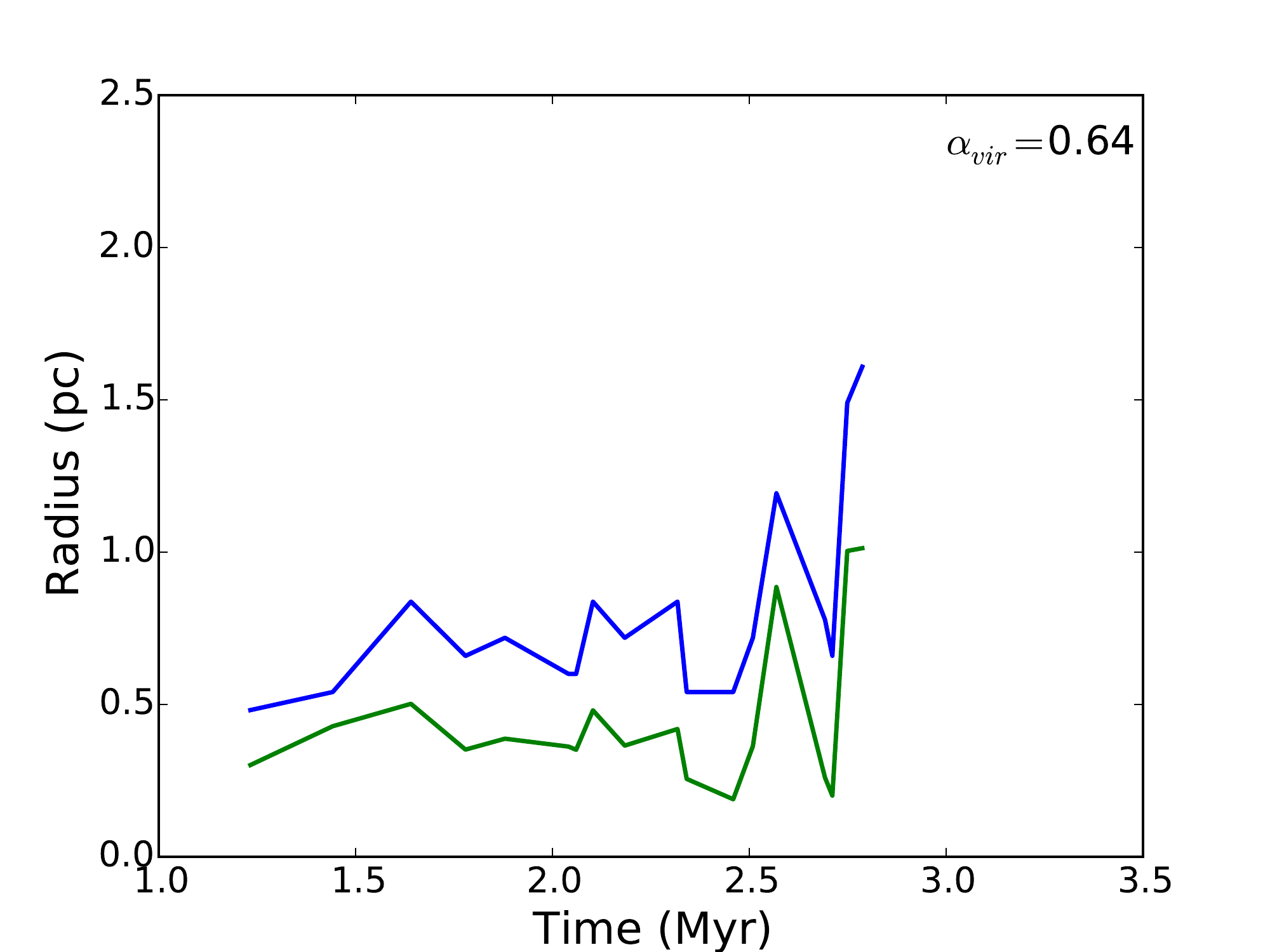}\caption{$(R^2H)^{1\over 3}, F_0$}
\end{subfigure}
\begin{subfigure}{.33\textwidth}
\includegraphics[trim=20 0 40 20,clip,width=\textwidth]{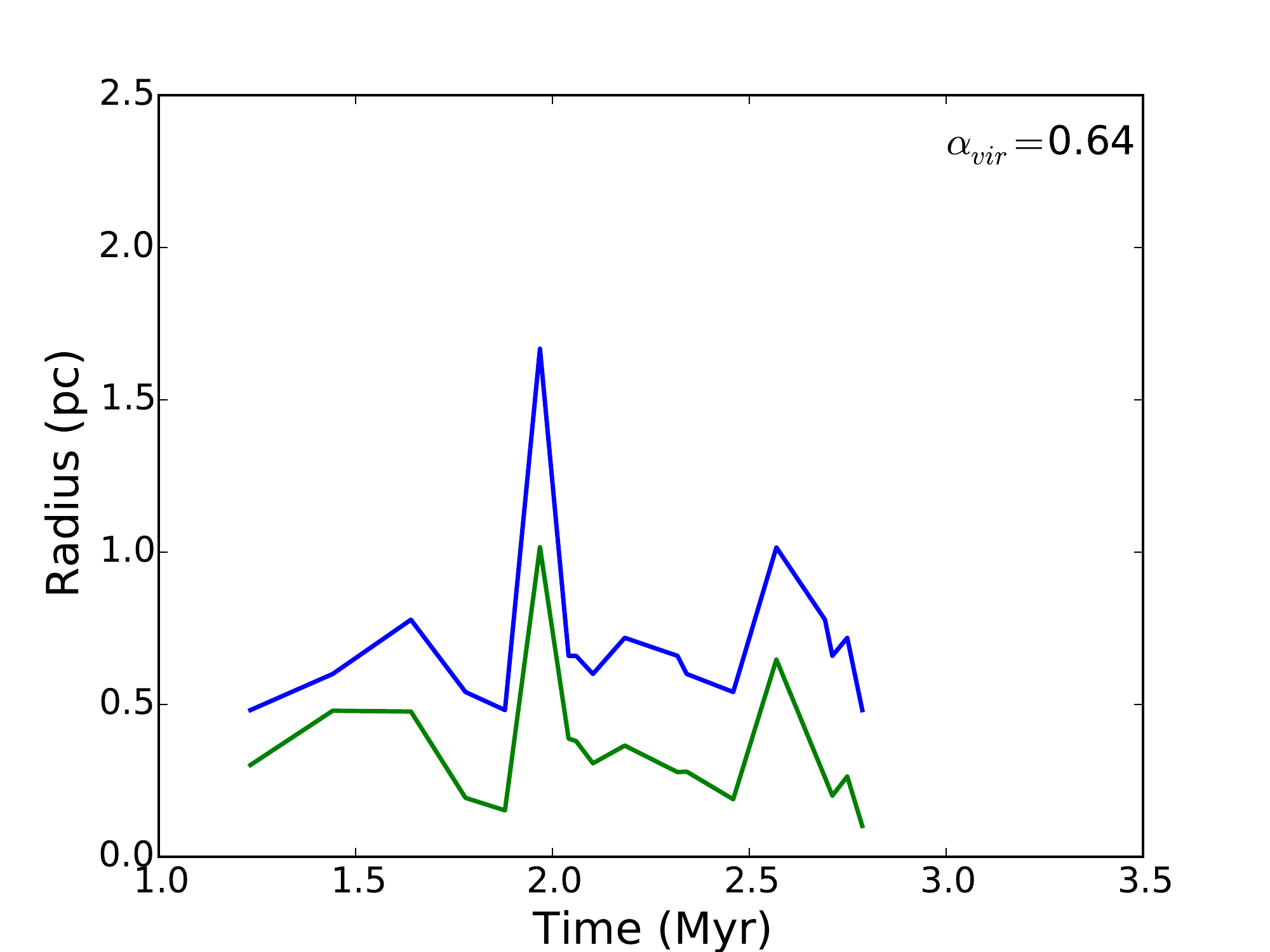}\caption{$(R^2H)^{1\over 3}, W$}
\end{subfigure}
\begin{subfigure}{.33\textwidth}
\includegraphics[trim=20 0 40 20,clip,width=\textwidth]{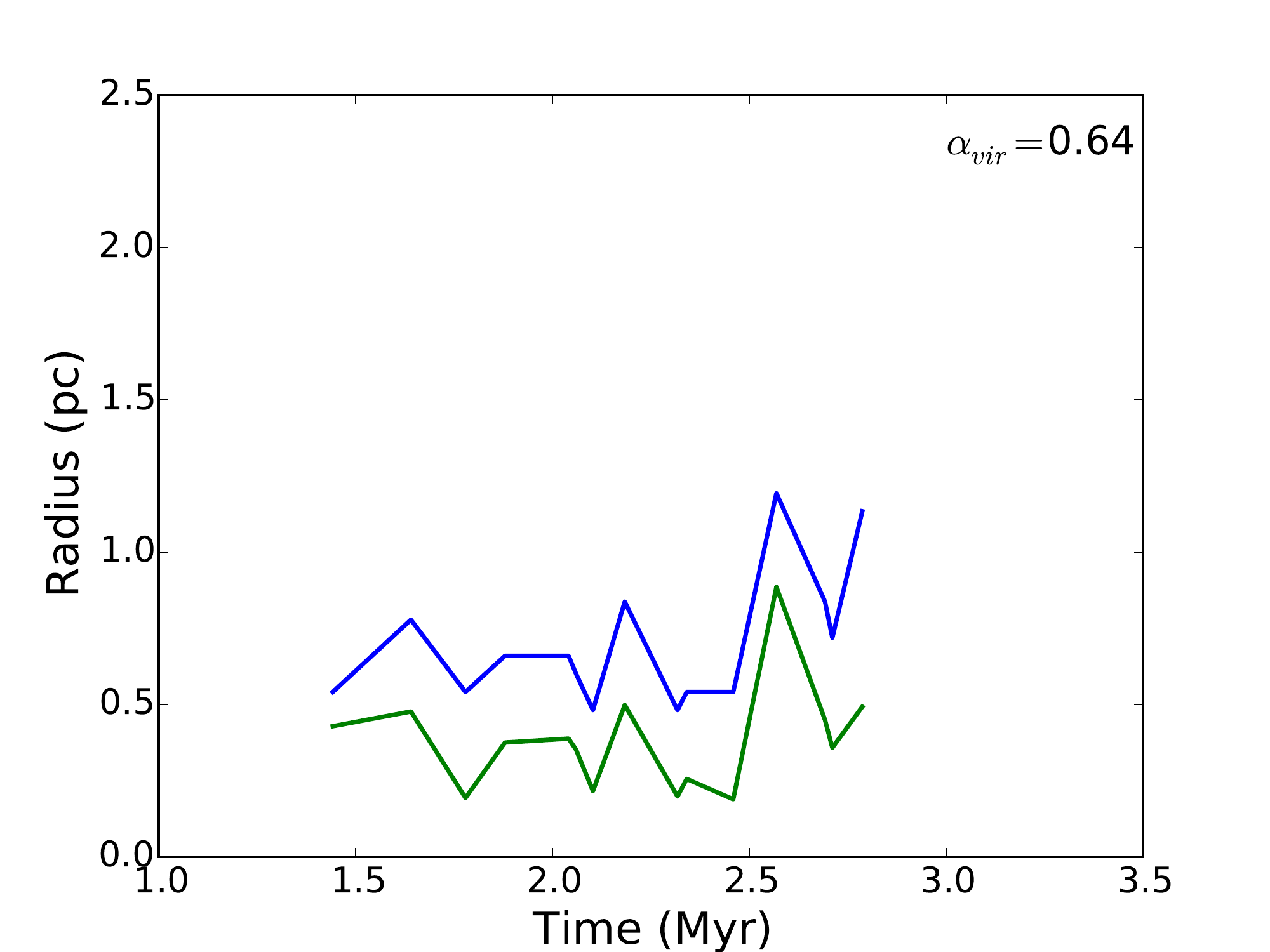}\caption{$(R^2H)^{1\over 3}, F$}
\end{subfigure}
\caption{Determination of gaseous proto-cluster size using various characteristic quantities for run B, same values as the left column of Fig. \ref{rt_mt_gas} are shown. The semi-major and semi-minor axes of the ellipsoidal proto-cluster is plotted against time. Panels (a), (b), and (c) use the infall over rotational velocity ($W_0$) versus weighted average radius $(R^2H)^{1\over 3}$, $(RH)^{1\over 2}$, and $R$ respectively. Panels (d), (e), and (f) use the geometrical averaged radius $(R^2H)^{1\over 3}$ to characterize the infall velocity ($F_0$), the distance weighted infall over rotation ($W$), and distance weighted infall ($F$) respectively. Not exactly same radius is discovered with different methods, but the averaged value is in general very close to one another.}
\label{rt_comps}
\end{figure*}

The gaseous proto-cluster is characterized by a transition from globally infalling motion to quasi-static rotational motion.
We use a piece-wise power-law fit to distinguish the envelope and the proto-cluster.
We are faced with the questions: 
\emph{1)} which quantity best represents the size of the proto-cluster? 
and \emph{2)} which quantity best characterizes the infall-rotation transition?
We choose the averaged radius  $R_\mathrm{avg}=(R^2H)^{1\over 3}$ 
and the mass weighted infall over rotational velocity $W_0 = \int \vec{v}\cdot \vec{n} ~dm / \|\int \vec{v}\times \vec{n}~dm\|$, 
where $\vec{v}$ is the velocity and $\vec{n}$ the unit vector pointing from the ellipsoid center to the cell position,
as characteristic quantities of the gaseous proto-cluster.
Here we show for one case (run B) the radius determined using $R$ or $\sqrt{RH}$ as the characteristic radius,
and also $F_0 = \int \vec{v}\cdot \vec{n} ~dm $, $W = \int \vec{v}\cdot \vec{r} ~dm / \|\int \vec{v}\times \vec{r}~dm\|$, and $F = \int \vec{v}\cdot \vec{r} ~dm $ 
, where $\vec{r}$ is the vector pointing from ellipsoid center to the cell, to characterize the transition of motion.
$F_0$ is the sum of infall velocity, 
$W$ is the distance-weighted infall over rotation measure, 
and $F$ is the distance-weighted infall moment.
We do not consider rotation alone since the rotational property does not show remarkable change when we cross the cluster border. 
The results of cluster radius using different parameters are compared in Fig. \ref{rt_comps}.
The level of fluctuations vary from one method to another, but they all catch similar proto-cluster size in general.
We conclude that our algorithm is able to capture the gaseous proto-cluster size robustly.

\section{Energy analysis}
\label{E_ana_others}
\begin{figure*}[]
\includegraphics[width=0.348\textwidth]{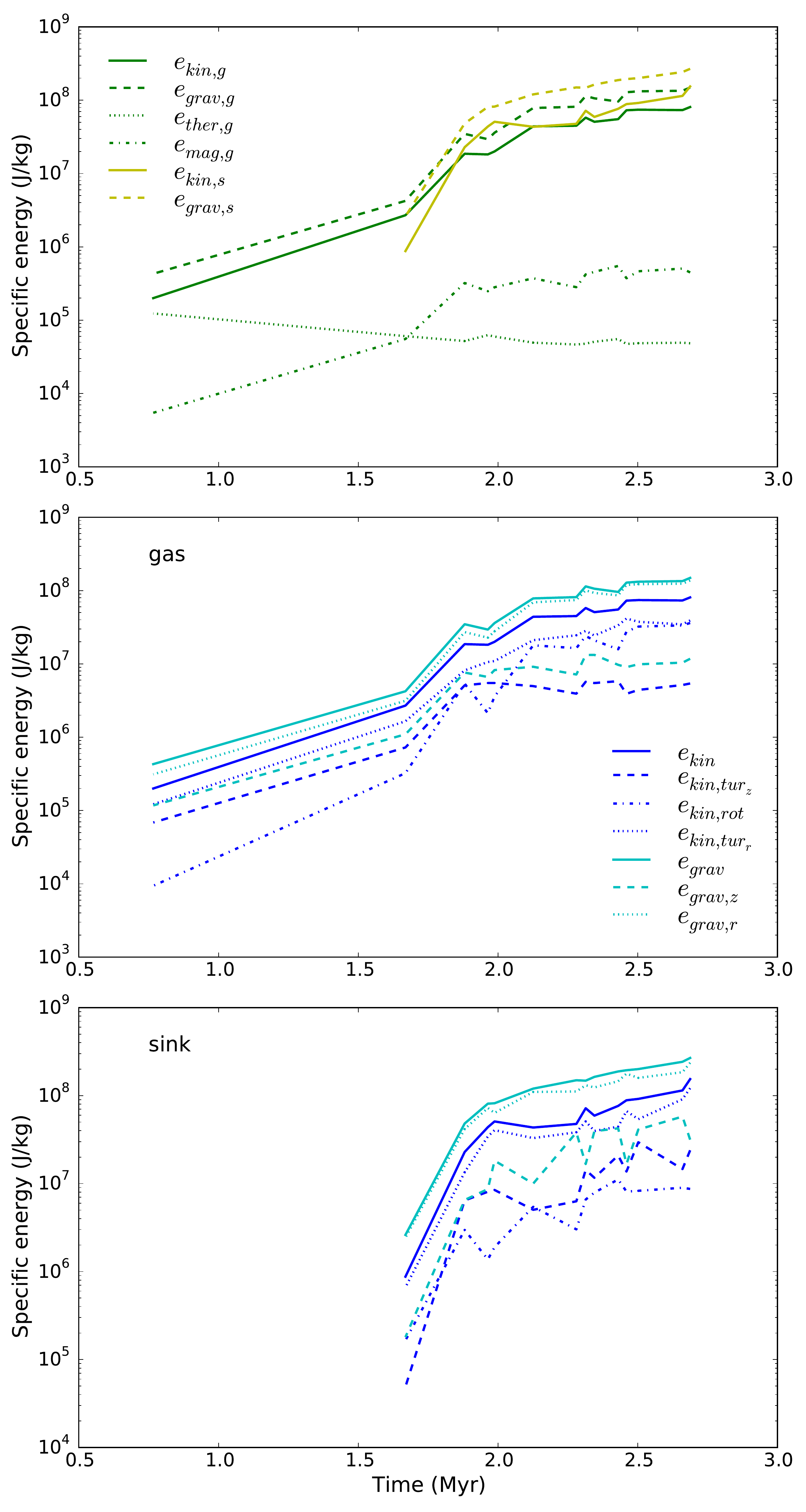}
\includegraphics[trim=35 0 0 0,clip,width=0.33\textwidth]{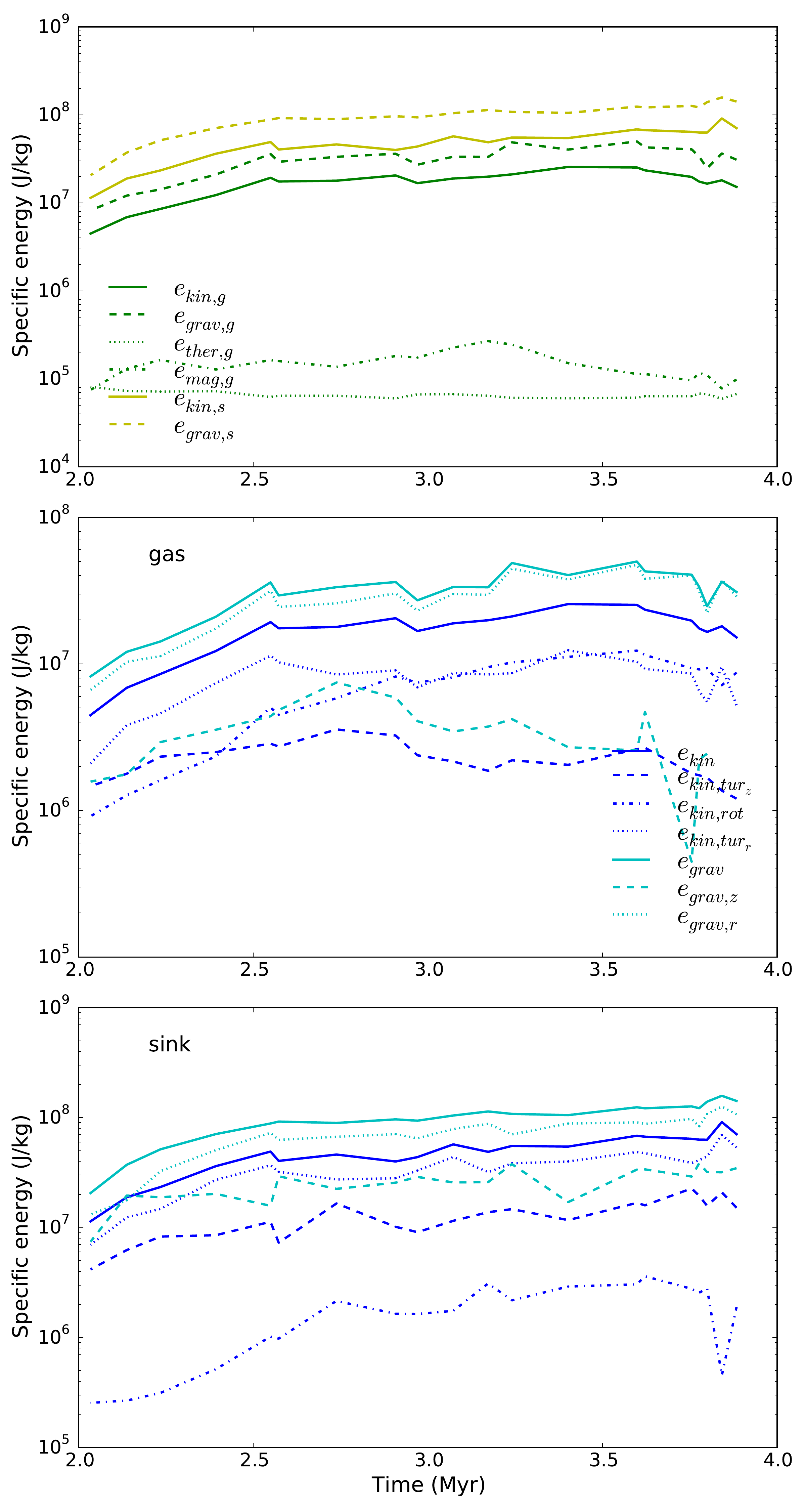}
\includegraphics[trim=35 0 0 0,clip,width=0.33\textwidth]{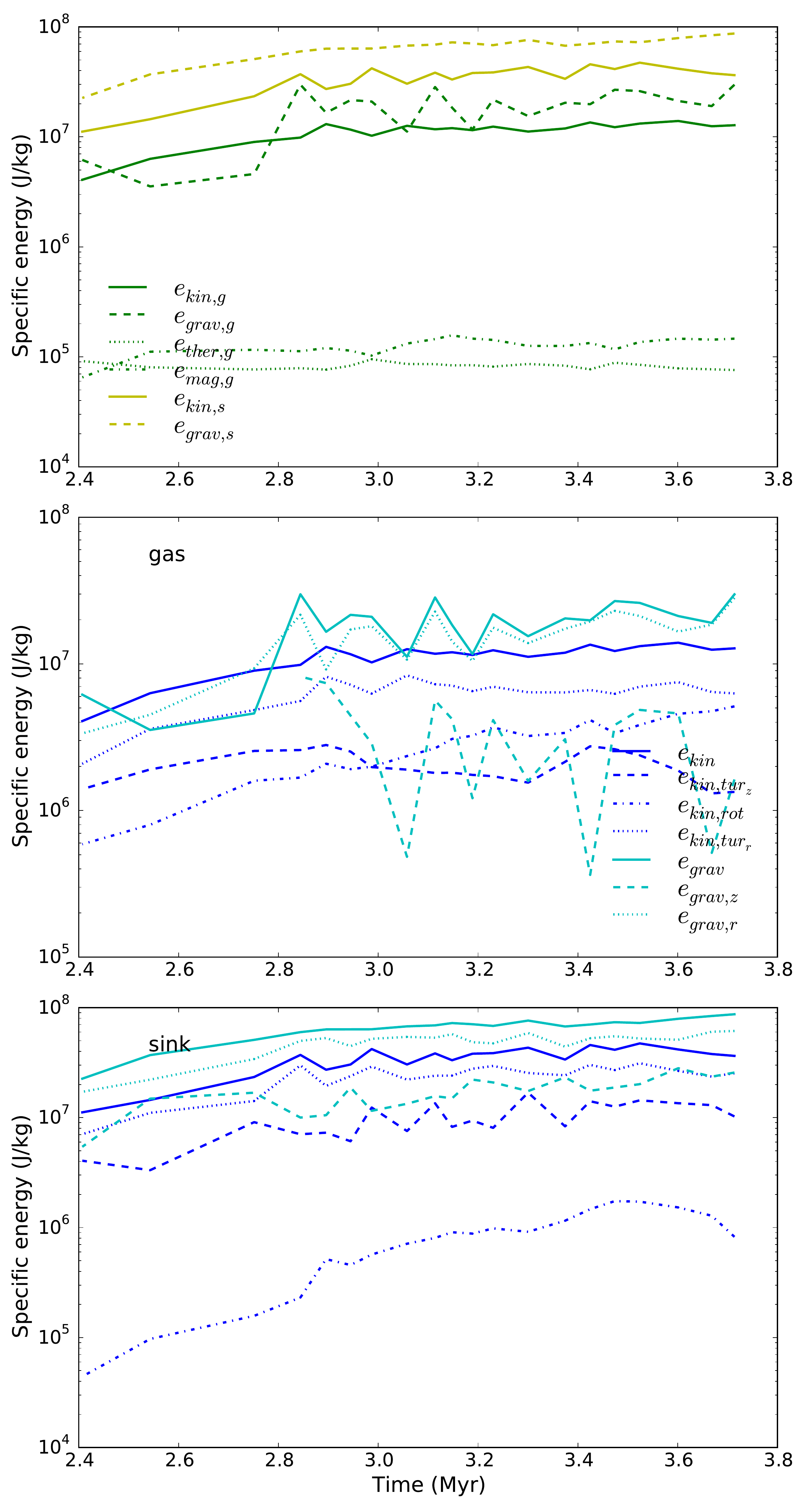}
\caption{Same plot as that in Fig. \ref{E_time} for other runs.
From left to right are runs with $t_\mathrm{ff}/t_\mathrm{vct} = 0.4, 1.1,~ \mathrm{and}~ 1.5$.
Top panel: specific energy in various forms of the gas and sink components.
Middle and bottom panels: dimensional energy analysis for gas and sinks respectively.
}
\label{E_others}
\end{figure*}
We show the same plot as that in Figs. \ref{E_time} for the three other runs with $t_\mathrm{ff}/t_\mathrm{vct} = 0.4, 1.1, ~\mathrm{and}~ 1.5$ in Fig \ref{E_others}. 
Similarly, virial equilibrium is observed for both gas and particle components,
and is valid even when the two dimension are analyzed separately.
The low turbulence case ($t_\mathrm{ff}/t_\mathrm{vct} = 0.4$) has little support at the beginning and is therefore prone to collapse, 
while its equilibrium is established at very early time.
The $t_\mathrm{ff}/t_\mathrm{vct} = 1.5$ case shows some fluctuations in gravitational energy and the proto-cluster is unbound at some moments, 
while this is possibly merely an artifact as we can see in Fig. \ref{rt_mt_gas} that sometimes a region larger than the proto-cluster is actually captured.


\bibliographystyle{aa}
\bibliography{biblio_cluster}

\end{document}